\documentclass[11pt,a4paper]{article}

\usepackage{jheppub}

\usepackage[utf8]{inputenc}
\usepackage[normalem]{ulem}
\usepackage{dcolumn}
\usepackage{array} 
\usepackage{bm}
\usepackage[dvipsnames]{xcolor}
\usepackage{graphicx}    
\usepackage{slashed}
\usepackage{multirow,bigstrut}
\usepackage{mathrsfs} 
\usepackage{relsize,amsmath}
\usepackage[geometry]{ifsym}
\usepackage{amssymb}

\usepackage{pifont}
\usepackage{multirow}

\usepackage{siunitx}

\usepackage{cleveref}

\hypersetup{%
    colorlinks = true,
    linktocpage = true,
    linkcolor =  blue,
    anchorcolor = purple,
    citecolor = magenta,
    filecolor = purple,
    urlcolor = BlueGreen
}

\graphicspath{{./figures/}}
\renewcommand{\arraystretch}{1.3}
\crefname{figure}{Figure}{Figures}
\Crefname{figure}{Figure}{Figures}
\crefname{section}{Section}{Sections}
\Crefname{section}{Section}{Sections}
\crefname{equation}{Eq.}{Eqs.}
\Crefname{equation}{Eq.}{Eqs.}
\crefname{table}{Table}{Tables}
\Crefname{table}{Table}{Tables}
\crefname{appendix}{Appendix}{Appendices}
\Crefname{appendix}{Appendix}{Appendices}

\newcommand*{\soft}{{3D}}
\newcommand*{\lamVL}{\ensuremath{\lambda_{A\Phi}}}
\newcommand*{\lamVLL}{\ensuremath{\lambda_{A}}}
\newcommand*{\muD}{\ensuremath{\mu_{D}}}

\title{\huge Beyond the Daisy Chain: Running and the 3D EFT View of Supercooled Phase Transitions} 

\author[a]{Martin Christiansen,}
\author[b]{Eric Madge,}
\author[a]{Cristina Puchades-Ib\'a\~nez,}
\author[a,c]{Maura E. Ramirez-Quezada,}
\author[a]{and Pedro Schwaller}

\affiliation[a]{PRISMA$^+$ Cluster of Excellence \& Mainz Institute for Theoretical Physics,\protect\\ Johannes Gutenberg University, 55099 Mainz, Germany}
\affiliation[b]{Instituto de F\'isica Te\'orica UAM/CSIC {\normalshape and} Departamento de F\'isica Te\'orica, Universidad Aut\'onoma de Madrid, Cantoblanco, 28049 Madrid, Spain}
\affiliation[c]{Dual CP Institute of High Energy Physics, C.P. 28045, Colima, M\'exico.}

\emailAdd{machrist@uni-mainz.de}
\emailAdd{eric.madgepimentel@uam.es}
\emailAdd{crpuchad@uni-mainz.de}
\emailAdd{mramirez@uni-mainz.de}
\emailAdd{pedro.schwaller@uni-mainz.de}

\subheader{\hspace*{\fill}{\tt IFT-UAM/CSIC-25-139}\newline\hspace*{\fill}{\tt MITP-25-071}}
\arxivnumber{2511.02910}
\abstract{%
    Pulsar timing arrays have recently observed a stochastic gravitational wave background at nano-Hertz frequencies. This raises the question whether the signal can be of primordial origin. Supercooled first-order phase transitions are among the few early Universe scenarios that can successfully explain it. To further scrutinise this possibility, a precise theoretical understanding of the dynamics of the phase transition is required. Here we perform such an analysis for a dark sector with an Abelian Higgs model in the conformal limit, which is known to admit large supercooling. 
    We compare simple analytic parametrisations of the bounce action, one-loop finite temperature calculations including Daisy resummation, and results of a dimensionally reduced (3D) effective theory including up to two-loop corrections using the DRalgo framework. Consistent renormalisation group evolution (RGE) of the couplings is essential for a meaningful interpretation of the results. We find that the 3D EFT with consistent expansion in the 4D parameters gives a significantly reduced scale dependence of the phase transition parameters. With a suitable choice of RGE scale, the 4D high temperature expanded effective potential yields results consistent with the 3D calculations, while the analytic parametrisation deviates significantly in the limit of large supercooling. 
}

\begin{document}

\maketitle


\section{Introduction}
\label{sec:introduction}

The discovery of gravitational waves (GWs) by the LIGO/Virgo Collaboration~\cite{LIGOScientific:2016aoc} has opened a new observational window into the Universe. 
GWs carry information not only about their astrophysical sources, but also about fundamental interactions in the early Universe, thus providing a unique complement to collider and astrophysical probes. While ground-based interferometers are sensitive to frequencies in the  $\mathcal{O}({\rm Hz}\text{--}{\rm kHz})$ range, pulsar timing arrays (PTAs) such as the North American Nanohertz Gravitational Wave Observatory (NANOGrav)~\cite{McLaughlin:2013ira,Brazier:2019mmu}, the European Pulsar Timing Array (EPTA)~\cite{Kramer:2013kea,Desvignes:2016yex},
the Parkes Pulsar Timing Array (PPTA)~\cite{Manchester:2012za}, the Indian Pulsar Timing Array (InPTA)~\cite{ChandraJoshi:2022etw}, the Chinese Pulsar Timing Array (CPTA)~\cite{Lee:2016xxx}, the MeerKAT Pulsar Timing Array (MPTA)~\cite{Miles:2022lkg}, and their combination, the International Pulsar Timing Array (IPTA)~\cite{Verbiest:2009kb,Manchester:2013ndt} target the nano-Hertz band, which is ideally suited to probe cosmological processes.

In their most recent data releases, all major PTAs report strong evidence for a stochastic gravitational-wave background (SGWB)~\cite{NANOGrav:2023gor,EPTA:2023fyk,Reardon:2023gzh,Xu:2023wog,Miles:2024seg} in this frequency range. While such a signal may originate from a population of supermassive black hole binaries~\cite{Rajagopal:1994zj,Jaffe:2002rt,Wyithe:2002ep,Sesana:2004sp,McWilliams:2012an,Burke-Spolaor:2018bvk,Becsy:2022pnr,Ellis:2023owy}, it might also point to a cosmological origin, such as a first-order phase transition (FOPT)~\cite{Kosowsky:1991ua,Schwaller:2015tja,Breitbach:2018ddu,Fairbairn:2019xog,Addazi:2020zcj,Nakai:2020oit,Neronov:2020qrl,Li:2020cjj,Borah:2021ftr,Morgante:2022zvc,Bringmann:2023opz,Megias:2023kiy,Addazi:2023jvg,Han:2023olf,NANOGrav:2021flc,Freese:2022qrl,Fujikura:2023lkn,Athron:2023mer,Jiang:2023qbm,Goncalves:2025uwh,Costa:2025csj}, cosmic defects like strings~\cite{Vachaspati:1984gt,Sakellariadou:1990ne,Ellis:2020ena,Blasi:2020mfx,Buchmuller:2020lbh,EuropeanPulsarTimingArray:2023lqe,Wang:2023len,Ellis:2023tsl,Bai:2023cqj,Kitajima:2023vre,Eichhorn:2023gat} and domain walls~\cite{Hiramatsu:2013qaa,Ferreira:2022zzo,Guo:2023hyp,Kitajima:2023cek,Gouttenoire:2023ftk,Lazarides:2023ksx,Blasi:2023sej}, inflation~\cite{Turner:1996ck,DeLuca:2020agl,Vagnozzi:2020gtf,Ashoorioon:2022raz,Vagnozzi:2023lwo,Cai:2023dls}, or other forms of new dynamics beyond the Standard Model (SM) in the early Universe~\cite{Ratzinger:2020koh,Bian:2020urb,Madge:2023dxc,NANOGrav:2023hvm,Franciolini:2023wjm,Ellis:2023oxs}.  Detecting such a background would provide direct information about high-energy physics far below the electroweak scale. 

In the SM, both the electroweak and QCD transitions are crossovers~\cite{Kajantie:1996mn,Aoki:2006we} and therefore do not produce an observable GW signal. 
However, in many extensions of the SM, spontaneous symmetry breaking can proceed via a FOPT, during which bubbles of the true vacuum nucleate and expand in the surrounding plasma. 
The resulting bubble collisions, together with the ensuing sound waves and turbulence, act as sources of a SGWB~\cite{Witten:1984rs,Hogan:1986dsh,Kamionkowski:1993fg,Ellis:2019oqb,Ellis:2020nnr,Lewicki:2022pdb,Kierkla:2022odc}. 
The characteristic shape and amplitude of the spectrum are determined by the thermodynamic parameters of the transition, which depend sensitively on the finite-temperature effective potential. Models featuring classical scale invariance offer a particularly appealing framework for strong FOPTs~\cite{Jinno:2016knw,Iso:2017uuu,Marzola:2017jzl,Azatov:2019png,Borah:2021ftr}. 
In such theories, symmetry breaking occurs radiatively via the Coleman-Weinberg mechanism~\cite{Coleman:1973jx}, naturally leading to significant supercooling and a substantial release of latent heat. 
The resulting transition can generate GW signals in the PTA frequency range, rendering these models directly testable against the NANOGrav observations~\cite{NANOGrav:2021flc,Freese:2022qrl,Madge:2023dxc,NANOGrav:2023hvm,Fujikura:2023lkn,Athron:2023mer,Jiang:2023qbm,Goncalves:2025uwh,Costa:2025csj,Balan:2025uke}.

A central theoretical challenge in predicting such GW signals lies in the precise computation of the finite-temperature effective potential. 
At finite temperature, infrared (IR) divergences emerge from Bose-enhanced low-energy bosonic modes (in particular the Matsubara zero modes), spoiling the perturbative expansion and introducing large uncertainties in the extraction of phase transition parameters~\cite{Linde:1978px,Linde:1980ts,Weinberg:1974hy,Dolan:1973qd}. Daisy resummation~\cite{Weinberg:1974hy,Dolan:1973qd,Parwani:1991gq,Arnold:1992rz} addresses this issue by resumming the leading IR-sensitive ring (Daisy) diagrams, which generate Debye thermal masses for soft bosonic modes and restore perturbative control. 
This technique remains the standard tool for estimating leading thermal corrections and provides a practical framework for studying cosmological phase transitions. However, Daisy resummation does not capture sub-leading yet relevant thermal effects, especially when multiple thermal scales are present or in strongly supercooled transitions, where the high-temperature expansion can break down~\cite{Gould:2021oba}.

Several methods beyond Daisy resummation have been developed to address this issue, including gap-equation-based dressing procedures and alternative resummation approaches~\cite{Boyd:1993tz,Curtin:2016urg,Curtin:2022ovx,Bahl:2024ykv,Bittar:2025lcr,Navarrete:2025yxy}, the heat-kernel expansion~\cite{Vassilevich:2003xt,Chakrabortty:2024wto,Balui:2025yvd}, as well as dimensional reduction~(DR)~\cite{Farakos:1994kx,Kajantie:1995dw,Braaten:1995cm}.
The dimensionally-reduced effective field theory (EFT) provides a systematic framework to resum thermal scales. 
By integrating out heavy ultraviolet modes, DR yields a three-dimensional EFT that describes the soft degrees of freedom governing the transition, consistently resumming all relevant thermal scales and improving perturbative convergence~\cite{Kajantie:1996mn,Ekstedt:2022bff,Gould:2021oba}. It further controls residual gauge dependence and IR pathologies that affect simpler treatments~\cite{Gould:2021ccf,Lofgren:2021ogg,Hirvonen:2021zej,Schicho:2022wty,Ekstedt:2022zro,Lofgren:2023sep,Gould:2023ovu}. In addition, the \texttt{DRalgo} package~\cite{Ekstedt:2022bff} has recently automated the matching procedure and the computation of the EFT parameters, making DR techniques accessible for generic BSM models.

A subtlety arises when applying DR to classically scale-invariant theories, where large field-to-temperature hierarchies place the system formally beyond the high-temperature expansion. Nevertheless, the method remains applicable once the relevant physical regimes are separated:  Daisy resummation and DR capture the thermal dynamics around the barrier, while the zero-temperature potential controls the broken-phase minimum~\cite{Kierkla:2023von}. 
Including renormalisation-group (RG) running of couplings ensures consistent matching across regimes and substantially reduces unphysical dependence on the renormalisation scale $\mu$. Together, these ingredients render the combined approach theoretically consistent even in the presence of large scale hierarchies. 

In this work, we explicitly address this issue by identifying an optimal choice for the renormalisation scale that minimises unphysical $\mu$-dependence and ensures consistency between the one-loop Daisy-resummed and the two-loop dimensionally-reduced effective potentials for a concrete benchmark model. In our analysis, we consider  a minimal classically scale-invariant model with a dark $U(1)_D$ gauge symmetry and a complex scalar charged under it. For strongly supercooled transitions, bubble collisions dominate the resulting GW spectrum~\cite{Bodeker:2009qy,Bodeker:2017cim,Ellis:2019oqb}, making an accurate determination of the phase-transition parameters crucial for reliable predictions. We systematically compare different treatments of the finite-temperature potential, including a simple analytic parametrization, the one-loop 4D high-temperature (HT) expansion with Daisy resummation and DR-based  potentials with one- and two-loop corrections.  We then study their impact on the predicted GW signal. 
A central focus of our analysis is the role of RG running: we use the two-loop potential to fix an appropriate renormalisation scale $\mu$ that minimises scale dependence and aligns the one-loop Daisy-resummed potential with the two-loop result.  We find that, once RG improvement is incorporated,  the 4D HT with Daisy resummation at this adequate scale yields predictions in very good agreement with the two-loop calculation in the scale-invariant benchmarks studied here.

We introduce the theoretical framework and the finite-temperature effective potentials in \cref{sec:Framework_Veff}, 
derive the phase-transition parameters and present the resulting GW spectra in \cref{sec:FOPT}, 
and conclude in \cref{sec:conclusions}.


\section{Theoretical Framework and Effective Potentials}
\label{sec:Framework_Veff}

Scenarios with nearly conformal dynamics undergoing a supercooled FOPT naturally involve substantial vacuum energy release~\cite{Konstandin:2011dr,Konstandin:2011ds,vonHarling:2017yew,Jinno:2016knw,Iso:2017uuu,Marzola:2017jzl,Marzo:2018nov,Prokopec:2018tnq,Baratella:2018pxi,Levi:2022bzt,Kierkla:2025vwp}. In this regime, the Universe remains trapped in the false vacuum well below the critical temperature, and when bubbles eventually nucleate, the resulting out-of-equilibrium dynamics can generate gravitational waves in the PTA frequency range.

A minimal and well-motivated realisation of radiative symmetry breaking is provided by the Coleman--Weinberg (CW) mechanism applied to an Abelian gauge theory~\cite{Coleman:1973jx}.  
We focus on a dark photon model consisting of a $U(1)_D$ gauge field and a complex scalar $\Phi = (\phi+i\chi)/\sqrt{2}$ carrying the dark-sector charge described by  
\begin{align}
    \label{eq:Lag}
    \mathcal{L} = 
    -\frac{1}{4} F_{\mu\nu}^2
    + \left|\left(\partial_\mu - i g A_\mu\right)\Phi\right|^2
    + m^2 |\Phi|^2
    - \lambda |\Phi|^4 \, .
\end{align}
At tree level we set $m^2 = 0$, ensuring that the potential is exactly scale-invariant at the classical level and contains no intrinsic mass scale.
Taking the limit $\lambda\rightarrow0$, 
radiative corrections then generate a non-trivial vacuum structure, spontaneously breaking scale invariance.

\begin{figure}
    \centering
    \includegraphics[width=0.49\linewidth]{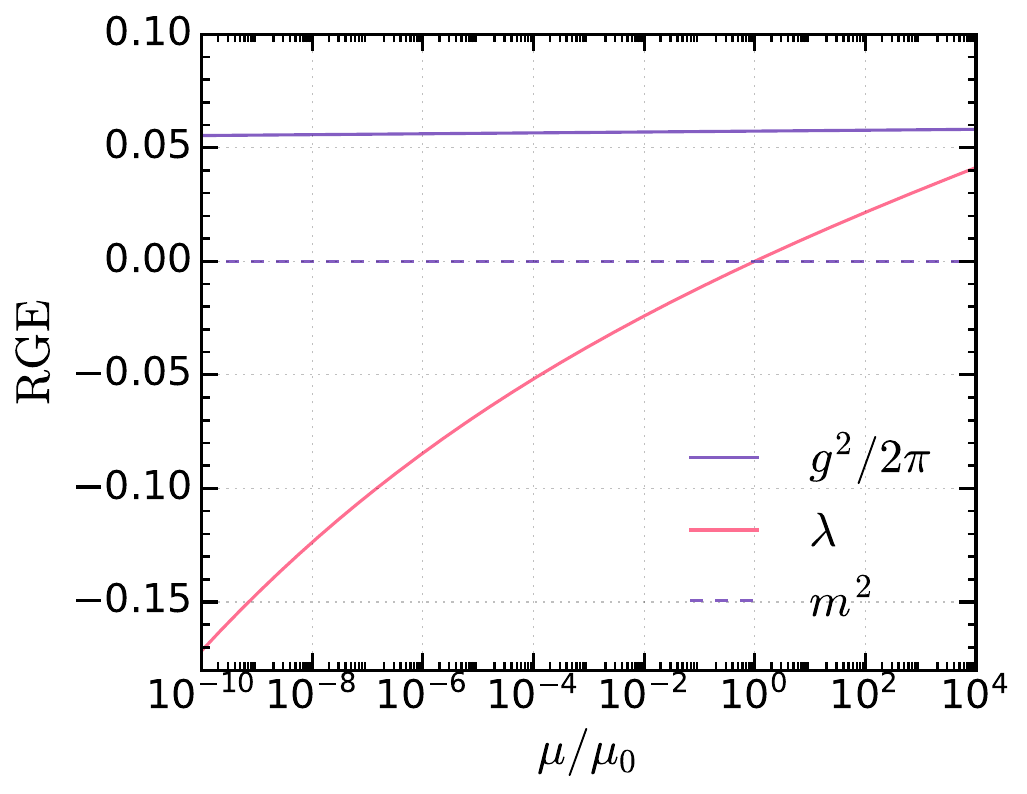}
    \includegraphics[width=0.49\linewidth]{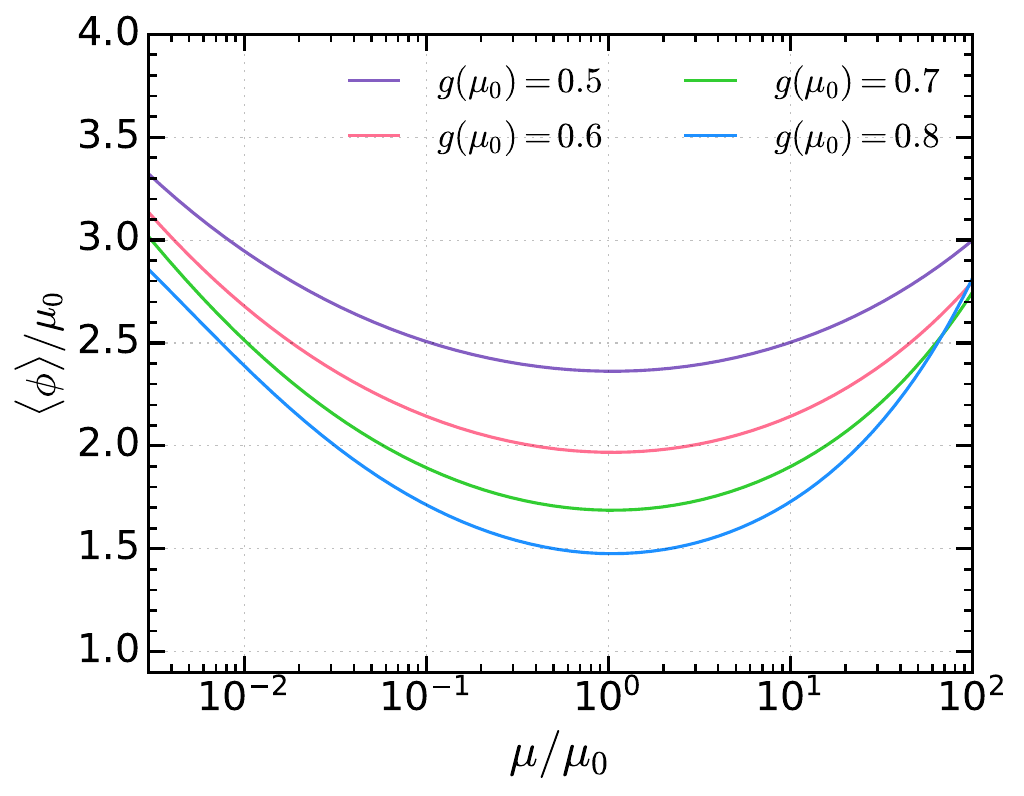}
    \caption{%
    \textbf{Left:} Renormalisation-group evolution of the model parameters for $g(\mu_0)=0.6$, 
    with input conditions $m^2(\mu_0)=0$ and $\lambda(\mu_0)=0$ at $\mu_0=\SI{1}{\GeV}$ 
    (see \cref{tab:mu_scales}). 
    The gauge coupling $g$ (shown as $g^2/2\pi$) runs slowly, 
    while the quartic coupling $\lambda$ decreases and becomes negative at intermediate scales, 
    triggering radiative symmetry breaking. 
    The mass parameter $m^2$ remains nearly constant, reflecting its vanishing initial condition. 
    The VEV emerges around the scale where $\lambda$ crosses zero, 
    marking the onset of spontaneous symmetry breaking. \textbf{Right:} The VEV as a function of $\mu$ for $g(\mu_0)=0.5-0.8$, $\lambda(\mu_0)=0$ and $m^2(\mu_0)=0$. The residual $\mu$-dependence is minimised for $\mu\sim\mu_0$, showing that the explicit scaling in \cref{eq:vev} is compensated once the running of the couplings is included.
}
    \label{fig:runningparam}
\end{figure}

A barrier between the false and true vacua appears once thermal corrections are included, reshaping the potential around the origin and inducing a first-order transition. A critical feature of such nearly conformal theories is the hierarchy of scales between the field value where the potential barrier forms and the position of the true minimum \cite{Kierkla:2023von}. 
This hierarchy originates from the flatness of the classical potential and the logarithmic nature of the CW corrections, which separate these scales exponentially.  
Because the effective potential probes field values across several orders of magnitude, large logarithms $\log(\phi/\mu)$ appear, and a proper RG improvement is essential to resum them and maintain perturbative control.  
In our model, $g$ and $m^2$ evolve slowly, while the quartic coupling $\lambda$ runs rapidly and turns negative at intermediate scales, setting the location and shape of the potential barrier. 
The scale where $\lambda$ crosses zero marks the onset of radiative symmetry breaking, as the potential develops a non-trivial minimum while $m^2$ remains small. 
This behaviour is illustrated in the left panel of \cref{fig:runningparam}, which shows the RG evolution (RGE) of the parameters.
The existence of such a hierarchy makes the choice of the renormalisation scale particularly delicate~\cite{ Croon:2020cgk}.  
In practice, the effective potential depends explicitly on the scale $\mu$ (cf.~\cref{eq:CW_potential}).  
While this dependence should cancel once all perturbative orders are included, at a fixed-order calculation,  an unphysical sensitivity to the renormalisation scale remains~\cite{Gould:2021oba}. As a consequence, varying $\mu$ within a reasonable range can noticeably shift the critical temperature and the strength of the transition, calling for a consistent RG  improved treatment and a physically motivated choice of the scale. This feature becomes particularly transparent in the scale-invariant limit of the model, where no intrinsic mass scale exists. At tree level, we set $m^2=0$ and take $\lambda(\mu_0)\simeq 0$ at a reference scale $\mu_0$, ensuring that the potential is classically flat while remaining non-vanishing. 

At one loop, the effective potential develops an explicit dependence on the renormalisation scale $\mu$, and its minimisation yields
\begin{equation}\label{eq:vev}
    \langle\phi\rangle =\frac{e^{1/6}\,\mu}{g}\equiv v ,
\end{equation}
thus identifying $\mu$ as the dynamically generated scale associated with dimensional transmutation. At first sight, this relation appears to fix the vacuum expectation value (VEV) through an arbitrary choice of $\mu$, seemingly reducing the freedom in setting the renormalisation scale.

This apparent inconsistency is resolved once the RG evolution of the couplings is consistently included. Since $g$, $\lambda$, and $m^2$ evolve with $\mu$, the explicit scale dependence of the potential is compensated by the implicit dependence carried by the running parameters. As a result, RG invariance is restored up to higher-order corrections, and the VEV becomes effectively independent of $\mu_0$, contrary to the naive expectation from~\cref{eq:vev}, as illustrated in the right  panel of \cref{fig:runningparam}.

In order to study the dynamics of the FOPT, we must construct the finite-temperature effective potential, which encodes the temperature-dependent vacuum structure and governs when and how the transition proceeds. 
At finite temperature, quantum fields can be expanded in Matsubara modes with discrete frequencies $\omega_n = 2\pi n T$~\cite{Matsubara:1955ws,Laine:2016hma}.
The non-zero modes acquire thermal masses of order $2\pi T$, making them heavy and well controlled within perturbation theory. 
In contrast, the bosonic zero modes remain light and dominate the long-distance (infrared) dynamics of the plasma. 
Loop diagrams involving these static modes receive enhanced infrared contributions that grow with the number of loops and eventually break down the naive perturbative expansion. 

In gauge theories, to obtain a consistent thermal description, one must resum the most infrared-sensitive diagrams, thereby incorporating the leading thermal corrections to the propagators of the light bosonic fields. 

In the following, we employ two complementary frameworks to address this issue:  
(i) the \textbf{Daisy (ring) resummation}, which captures the leading plasma screening effects by reorganising the perturbative series at one loop, and  
(ii) \textbf{dimensional reduction (DR)}, which integrates out the heavy non-zero Matsubara modes, yielding an effective three-dimensional theory in which these resummations are automatically encoded and higher-order corrections can be treated systematically~\cite{Chala:2024xll,Bernardo:2025vkz,Chala:2025oul,Chala:2025cya}. The DR framework presents results less sensitive to gauge choices.~\cite{Gould:2021ccf,Lofgren:2021ogg,Hirvonen:2021zej,Schicho:2022wty,Ekstedt:2022zro,Lofgren:2023sep,Gould:2023ovu}.

The Daisy resummation approach directly resums a class of ring diagrams in four-dimensional finite-temperature perturbation theory by replacing the field-dependent masses with their thermal counterparts. Using the Arnold-Espinosa prescription, in the Landau gauge $\xi=0$, the Daisy contribution reads~\cite{Arnold:1992rz}\footnote{We have explicitly verified that, for the one-loop high-temperature potential of the CW model, using the Parwani prescription~\cite{Parwani:1991gq} or omitting Daisy resummation altogether produces qualitatively similar results to those obtained with the Arnold-Espinosa implementation~\cite{Arnold:1992rz} in the parameter region of interest.
}  
        \begin{equation}
        V_{\rm Daisy}(\phi,T)= - \sum_i \frac{n_i T}{12\pi} 
        \biggl[ 
        \bigl(m^2(\phi) + \Pi(T))_i^{3/2} -(m_i^2(\phi)\bigr)^{3/2} 
        \biggr] ,
        \label{eq:VDaisy}
    \end{equation}
where $n_i$ are the degrees of freedom, $m_i^2(\phi)$ the field-dependent masses (see \cref{eq:masses}), and $\Pi_i(T)$ the thermal masses (see \cref{eq:thermalmasses}). 
The subtraction avoids double counting. This method, however, is difficult to implement beyond one loop order, and does not reduce the residual scale dependence on $\mu$ at one loop.
By contrast, DR integrates out the heavy (non-zero) Matsubara modes, yielding a three-dimensional (3D) effective field theory for the soft zero modes.  
We follow Ref.~\cite{Kierkla:2023von} and work in the soft theory including both spatial\footnote{Note that spatial gauge bosons act as scale-shifters and require special attention, in particular when attempting to calculate the prefactor of the nucleation rate~\cite{Gould:2021ccf,Kierkla:2025qyz}.} and temporal Matsubara-zero modes of the gauge fields.
This procedure automatically incorporates Daisy resummation and provides a framework for a systematic loop expansion.
We employ the \texttt{DRalgo} package~\cite{Ekstedt:2022bff} to perform the matching of couplings and masses, run the RGEs, and compute the two-loop effective potential in three dimensions. 
Although DR at two loops is our best available approximation, it is still subject to truncation uncertainties in the matching procedure, which should be kept in mind.

In practice, these two resummation strategies lead us to consider three classes of effective potentials for the phase-transition analysis: the four dimensional (4D) one-loop potentials, the one-parameter approximation~(OPA)~\cite{Levi:2022bzt} and the 3D potentials from DR. We describe each potential in the following.

\paragraph{4D one-loop potentials.}
    
These include the Coleman-Weinberg term, finite-temperature corrections, and Daisy resummation.  
We distinguish between the 4D “Full” potential, where the thermal function $J_b(m_i^2/T^2)$ is evaluated numerically (see \cref{eq:JbInt}), and the 4D “High-Temperature” (HT) expansion potential, where $J_b$ is expanded in $m_i^2/T^2$, neglecting terms of order $\mathcal{O}(m_i^6/T^6)$ and higher.  

\paragraph{One-parameter approximation.}  
    
In addition to the Full and HT potentials, we also consider OPA~\cite{Levi:2022bzt}, a simplified 4D parametrisation proposed for supercooled scenarios.
In this parametrisation we neglect Daisy resummation, but it is particularly convenient for numerical studies, since it eliminates numerical instabilities and speeds up parameter scans.
In the HT limit, the effective potential can be approximated as a fourth-order polynomial in the field. In the calculation of the tunnelling action, the potential can then be rescaled to depend on a single parameter, and the action can be calculated from a one-dimensional function of this parameter.
The OPA is therefore equivalent to the 4D HT potential without Daisy corrections.

\paragraph{3D potentials from dimensional reduction.}     
After integrating out the non-zero Matsubara modes, the EFT parameters depend on the loop order used in the matching (see \cref{app:VeffDR}). We study four schemes:
    \begin{enumerate}
        \item \textbf{3D 1-L (LO).} One-loop potential with LO masses and couplings.
        \item \textbf{3D 1-L (NLO).} One-loop potential with NLO masses and couplings.
        \item \textbf{3D 2-L (NLO).} Two-loop potential with NLO masses and couplings.
        \item\textbf{3D 2-L (Mixed).} Two-loop potential with mixed input (tree level at NLO, higher orders at LO).
    \end{enumerate}
Schemes {3D 1-L (LO)} and {3D 2-L (Mixed)} correspond to fixed-order expansions in the 4D parameters $g^2$, $\lambda$ and $m^2$, whereas {3D 1-L (NLO)} and {3D 2-L (NLO)} partially include higher orders.
Note however that this power-counting assumes $\lambda \sim g^2$ and therefore is, strictly speaking, not valid in the supercooled regime where $\lambda \lesssim g^4$. In all schemes, we proceed as follows: we define the 4D couplings at $\mu_0$ and evolve them using the 4D RGEs to the matching scale $\mu_{\rm Match}$, where the 4D-to-3D dimensional-reduction matching is performed. The 3D effective potential is then evaluated using the soft scale choice $\mu_3=gT$, which enters through logarithms in the matching coefficients and in the 3D NLO potential. The scales used in each scheme are summarised in \cref{tab:mu_scales}.
\begin{table}
\centering
\renewcommand{\arraystretch}{1.2}
\setlength{\tabcolsep}{8pt}
\begin{tabular}{llccc}
\hline
 Approach & Potential & \multicolumn{3}{c}{Scales}\\
\hline
\textbf{4D}   & HT &  \multicolumn{3}{c}{$\mu=\pi T$}  \\[2pt]
\multirow{4}{*}{\textbf{DR 3D}}
              & 1--L (LO)   &
              \multirow{4}{*}{$\left.\begin{array}{c}\\ \\ \\ \\ \end{array}\right\}\ \ \mu_{\rm Match}=2\pi T$}
              &                             & \\[2pt]
              & 1--L (NLO)  &              &
              \multirow{3}{*}{$\left.\begin{array}{c}\\ \\ \\ \end{array}\right\}\ \ \mu_{3}=gT$}
              & \\[2pt]
              & 2--L (NLO)  &              & & \\[2pt]
              & 2--L (Mixed)&              & & \\
\hline
\multicolumn{5}{c}{Input: $\mu_0=1~\rm GeV \quad m(\mu_0)^2=0 \quad \lambda(\mu_0)=0 \quad g(\mu_0)\in[0.5,1]$}\\
\hline
\end{tabular}
\caption{
Summary of the renormalisation and matching scales used in the 4D and 3D DR effective potentials. 
The 4D HT potential is evaluated at the  renormalisation scale $\mu=\pi T$. 
In the 3D effective theory, $\mu_{\rm Match}$ denotes the matching scale at which heavy Matsubara modes are integrated out, 
and $\mu_3$ the soft scale where the three-dimensional potential is computed. 
Input parameters are defined at $\mu_0 = 1~\mathrm{GeV}$.
}
\label{tab:mu_scales}
\end{table}

\begin{figure}
    \centering
     \includegraphics[width=0.49\textwidth]{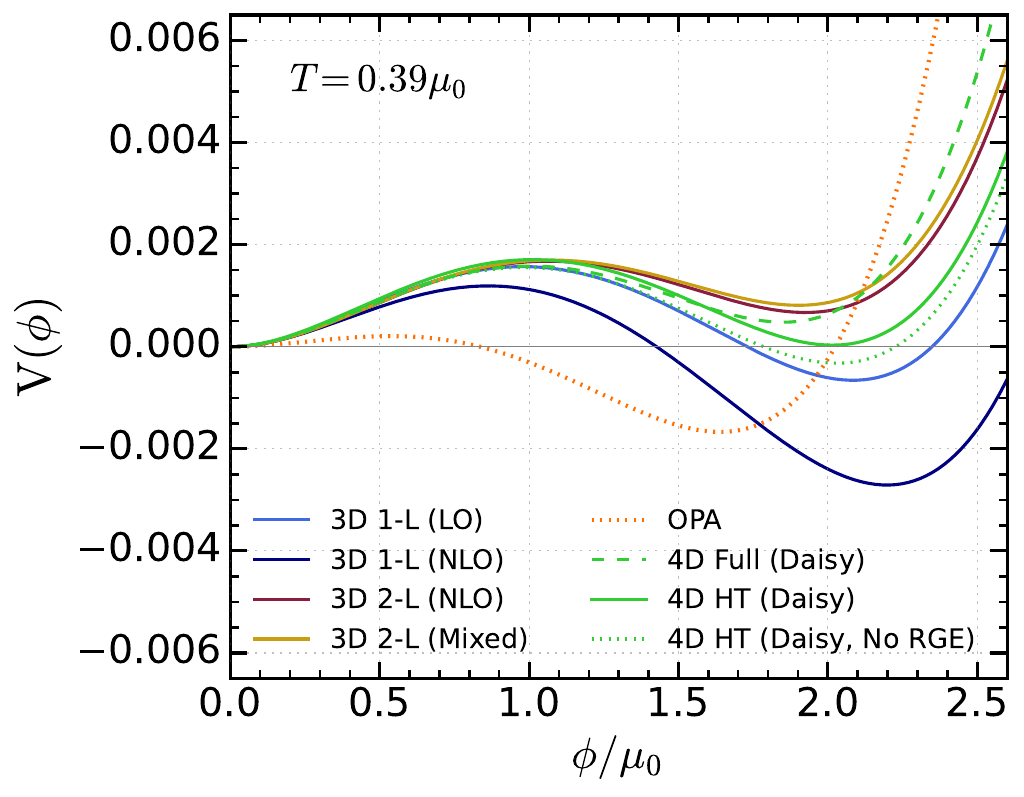}
     \includegraphics[width=0.49\textwidth]{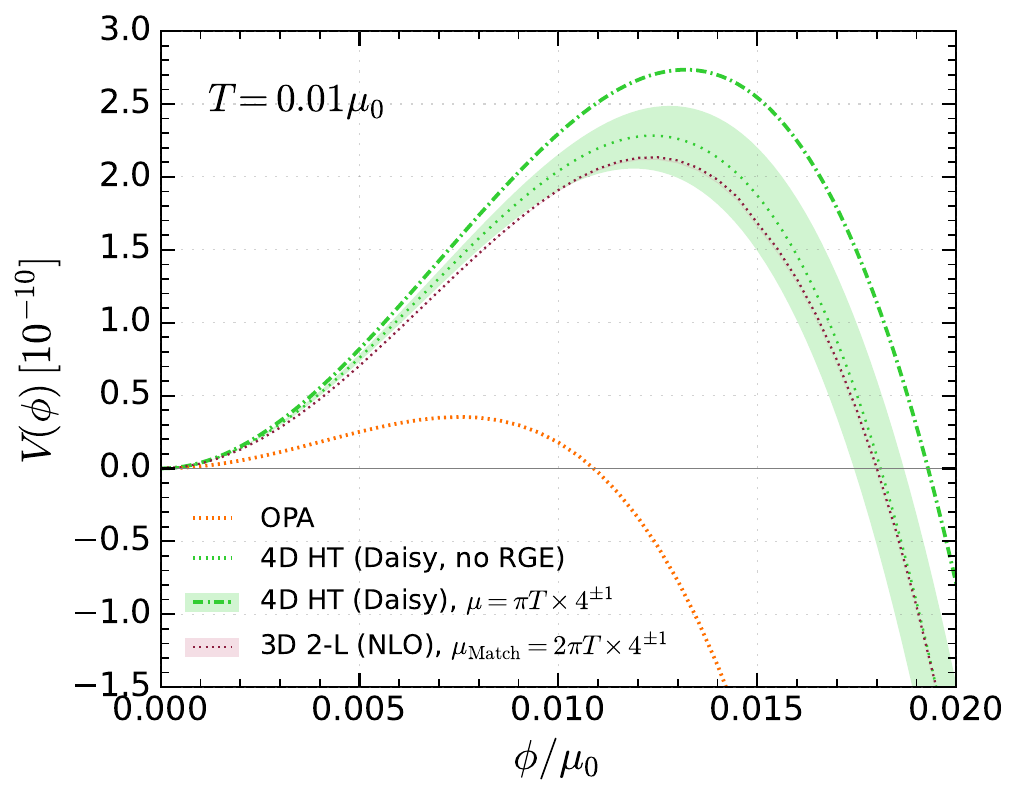}
\caption{Comparison of different prescriptions for the effective potential at $g(\mu_0)=0.6$. 
\textbf{Left:} Comparison between the 4D Full potential, its  4D HT expansion, the OPA, and the 3D dimensionally reduced potentials at different loop orders at $T=0.39\, \mu_0$.  
\textbf{Right:} Residual dependence on the renormalisation scale at $T = 0.01\, \mu_0$. 
For the 4D HT potential (green band), the renormalisation scale is set to $\mu = \pi T$ and varied between $\mu = \pi T/4$ and $\mu = 4\pi T$ to illustrate the scale dependence. 
For the 3D two-loop (NLO) potential (purple band), the scale is instead fixed at $\mu_{\rm Match} = 2\pi T$ and varied analogously from $\mu_{\rm Match} = (2\pi T)/4$ to $\mu_{\rm Match} = 4(2\pi T)$.
In each case, the dotted line shows the potential evaluated at the corresponding reference scale used in the corresponding approach. The dash-dotted green line corresponds to the 4D~HT potential at $\mu=\mu_0$, whereas the orange dotted line depicts the OPA (without running).}
 \label{fig:Veff_comp}
\end{figure}

The left panel of \cref{fig:Veff_comp} contrasts the 4D Full and HT prescriptions with the different DR schemes, illustrating how they affect the shape of the effective potential. 
To compare the 4D and 3D EFTs, we map the 3D potential to 4D multiplying it by the temperature $T$ and rescaling the field $\phi \to \phi/\sqrt{T}$, enabling a direct qualitative comparison between the two frameworks.
While, for field-independent renormalisation scale choices, RG running can be incorporated in the OPA (see \cref{app:OPA}), we here use the original form in Ref.~\cite{Levi:2022bzt} and use it as a benchmark to illustrate the effect of neglecting the RG running.
Hence, 
the OPA is computed by fixing the renormalisation scale to $\mu_0$ and the parameters to their tree-level values (last row of \cref{tab:mu_scales}), without including RG running, and neglecting Daisy resummation. 
As a result, it shows a significant deviation from the other prescriptions. This trend persists in the phase transition parameters discussed in the next section.  
This underscores the importance of properly accounting for RG running, as neglecting it can lead to a systematic underestimation of critical quantities.
The agreement between the 3D and 4D approaches after including RG effects further improves at lower temperatures, in particular in the region around the barrier (cf.\ right panel of \cref{fig:Veff_comp})

Among the approaches that incorporate RG running, the agreement between the 4D and 3D schemes is noticeably improved, with the thermal one-loop potential~(green) exhibiting even better agreement with the DR two-loop potential~(brown/gold) than DR at the one-loop level~(blue).
The largest deviation appears for the 3D one-loop (NLO) potential, exceeding even the difference between the 4D HT and the two-loop (NLO/Mixed) results.  
We interpret this as a consequence of introducing thermal masses and running couplings at NLO while the potential itself remains at  one-loop, which enhances  the residual scale dependence in both the 4D matching scale  $\mu_{\rm Match}$  and  the soft scale $\mu_3$, in the absence of compensating higher-order terms that would stabilise it.  
This highlights an important point: partial higher-order corrections do not necessarily improve the accuracy.
Incomplete higher-order inputs can degrade reliability and yield larger deviations rather than systematic improvement. 

In scale-invariant models this residual dependence is particularly pronounced; even with thermal and Daisy resummations, the 4D HT one-loop potential remains noticeably $\mu$-sensitive~\cite{Gould:2021oba}. To illustrate this, the right panel of \cref{fig:Veff_comp} explicitly displays the residual renormalisation-scale dependence of the finite-temperature potential at $T=0.01\,v$. In the 4D HT case (green band) we vary $\mu\in[\pi T/4,\,4\pi T]$ about the reference choice $\mu=\pi T$, and in the 3D two-loop (NLO) case (purple band) we vary $\mu_{\rm Match}\in[(2\pi T)/4,\,4(2\pi T)]$ about $\mu_{\rm Match}=2\pi T$. These bands should not be interpreted as uncertainty estimates; they simply quantify the $\mu$-dependence that remains at fixed perturbative order.
Varying $\mu$ manifests as a shift in the overall curvature, with visible changes in the barrier height and in the positions of the minima, arising from the explicit $\log(m_i^2/\mu^2)$ terms in the Coleman-Weinberg contribution and their finite-$T$ counterparts. In the 4D HT setup, where only Debye masses are resummed, the residual renormalisation-scale dependence is comparatively larger, which explains the wider band. By contrast, in the 3D two-loop (NLO) DR result, matching and running of the 3D couplings substantially reduce the residual scale dependence, leading to a much narrower band.

We exploit this behaviour to minimise the unphysical $\mu$-sensitivity when extracting the phase-transition parameters. 
Concretely, we take the 3D two-loop NLO potential as a benchmark and fix the 4D HT renormalisation scale by requiring the good agreement with the bounce action, which is presented in the following section. 
This criterion sets the renormalisation scale to $\mu=\pi T$ in our 4D HT calculations; we use this as our reference hereafter. We have explicitly checked that including the leading dimension-six operators in the 3D finite-temperature potential~\cite{Chala:2024xll,Bernardo:2025vkz} does not noticeably modify the barrier in the temperature range around $T_n$ that controls the bounce action, so their effect on the bounce action $S_3$ is negligible for our purposes.


\section{Phase-Transition Parameters and Inputs to the GW Spectrum}
\label{sec:FOPT}

The prediction of a SGWB signal from a FOPT is determined by a set of macroscopic parameters: the transition temperature $T_*$, the strength parameter $\alpha$, the inverse duration $\beta/H$, and the wall velocity $v_w$ (see e.g.\ Ref.~\cite{Athron:2023xlk} for a comprehensive review).  
In this work we set $v_w=1$, corresponding to the relativistic limit. While smaller wall velocities can affect the relative contribution of the different sources, they do not qualitatively modify our main conclusions for strong transitions.
All of these macroscopic quantities are determined through the thermal tunnelling action $S_3/T$.
Therefore, obtaining a reliable evaluation of $S_3(T)/T$ is essential, since it is the central object from which the transition parameters are extracted. 

\subsection{Bounce Action: Method and Validation}

We compute $S_3(T)$ using the semi-analytical method proposed by Espinosa~\cite{Espinosa:2018hue}, which estimates the bounce action from a fourth-order polynomial approximation to the tunnelling potential. 
This approach offers a substantial computational speed-up compared to numerical solutions of the corresponding differential equation, while exhibiting only minor deviations from the exact result in the parameter range of interest.
For validation, we have also compared our results with those obtained from a direct shooting method that numerically solves the bounce equation. The results show good agreement, confirming the reliability of the semi-analytical approach.

\begin{figure}
        \centering
        \includegraphics[width=0.49\linewidth]{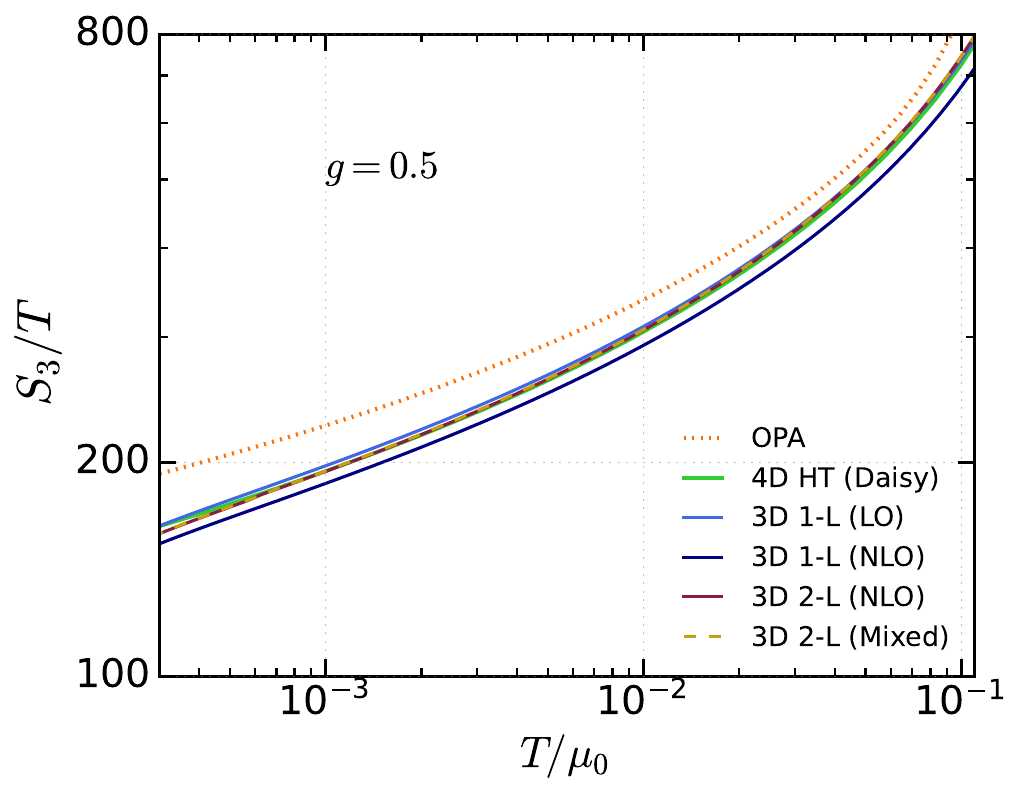}
        \includegraphics[width=0.5\linewidth]{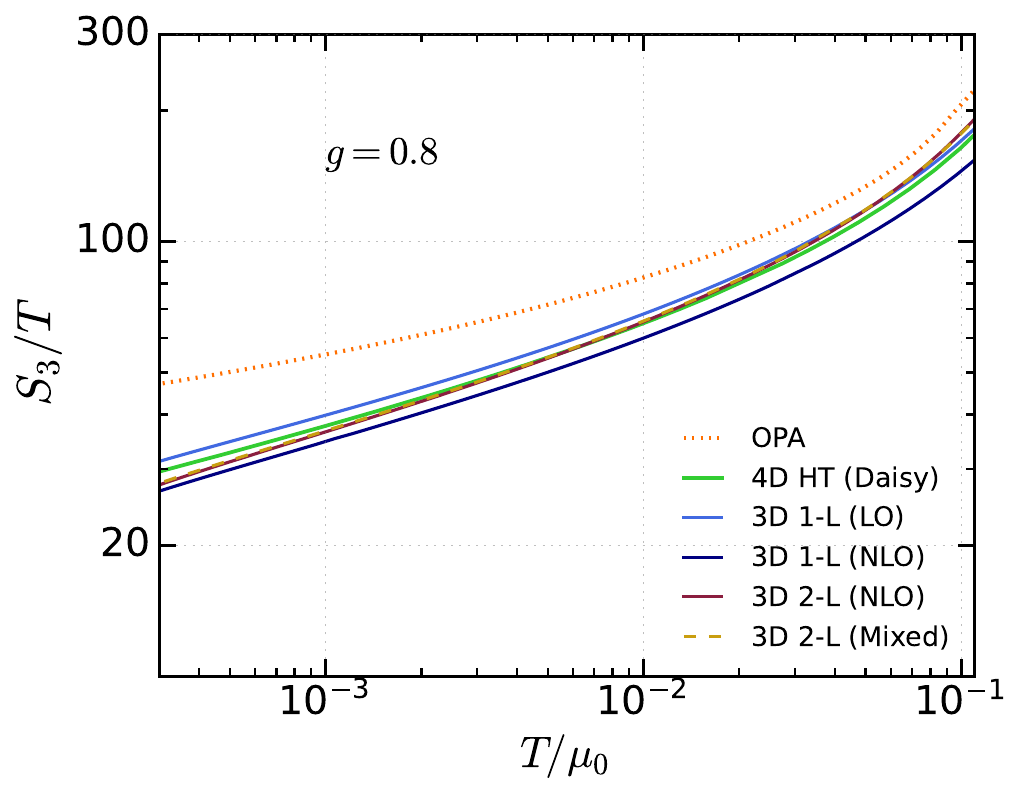}
        \caption{Comparison of the tunnelling action $S_3(T)/T$ computed with different effective potential schemes: 4D HT, one-parameter approximation (OPA), and dimensional reduction (DRalgo). Results are shown for $\mu_0=1$ and $\lambda(\mu_0) = 0$, at two benchmark values of the dark gauge coupling. 
        \textbf{Left:} $g(\mu_0)=0.5$, where all schemes are in close agreement. 
        \textbf{Right:} $g(\mu_0)=0.8$, where discrepancies between methods become more pronounced due to the increased sensitivity to thermal corrections at larger coupling.
        }
        \label{fig:S3_comparison}
    \end{figure}

In \cref{fig:S3_comparison} we show  $S_3(T)/T$ computed with different thermal resummation schemes for two representative values of the gauge coupling, $g=0.5$ and $g=0.8$. In general, schemes based on DR agree very well, particularly the two-loop approaches. Evaluating the HT potential at the optimal scale $\mu=\pi T$, as discussed in the previous section, also yields results comparable to the two-loop DR scheme, which we consider the most accurate. In contrast, OPA, which  is shown without running effects, deviates significantly. This highlights the importance of coupling running. As shown in the right panel of \cref{fig:S3_comparison}, the discrepancies between the different approaches become more pronounced at larger couplings. This is expected, since stronger couplings enhance thermal corrections and modify the shape of the potential barrier, increasing the sensitivity of the tunnelling action to higher-order effects and to the details of the resummation scheme. 
    
\subsection{Macroscopic Parameters}
The tunnelling action $S_3(T)$ governs the thermal nucleation rate per unit volume,
    \begin{equation}
    \Gamma(T) \simeq T^4\left(\frac{S_3(T)}{2\pi T}\right)^{3/2} e^{-S_3(T)/T}.
    \label{eq:nuc_rate}
    \end{equation}

Note that a more systematic treatment of the nucleation rate would require a perturbative expansion of $S_3$ and a corresponding evaluation of the fluctuation determinant~\cite{Hirvonen:2021zej}.
We leave such an analysis to future work.
At the level of precision required to explore the parameter space of interest, the neglected higher-order corrections are not expected to qualitatively affect our results.

From here, the nucleation temperature $T_n$ is defined by the condition
    \begin{equation}
        \Gamma(T_n) \simeq H^4(T_n),
        \label{eq:nuc_cond}
    \end{equation}
corresponding to the temperature at which approximately one critical bubble nucleates per Hubble volume.
The Hubble parameter during the radiation-dominated era is given by
\begin{equation}
  H^2(T,g;\mu)= \frac{8\pi G}{3}\left[\Delta V_{\rm eff}^0(g;\mu) + \rho_R(T)\right] \, ,
  \label{eq:Hubble}
\end{equation}
where  $\Delta V_{\rm eff}^0(g;\mu) = V_{\rm eff}^0(0,g;\mu)
    -V_{\rm eff}^0(v,g;\mu) = | V_{\rm eff}^0(v,g;\mu)|$ denotes the difference between the false and true vacuum of the zero-temperature potential, $\rho_R(T)=\pi^2g_*T^4/30$ is the radiation energy density, $G$ is the gravitational constant, and $g_*$ denotes the number of relativistic degrees of freedom.
While the percolation temperature $T_p$ gives a more accurate estimate of the transition temperature, in particular for supercooled PTs, where it can differ substantially from $T_n$~\cite{Athron:2022mmm}, we here adopt $T_* = T_n$. This choice is motivated by its computational simplicity and is sufficient for our purpose of conducting a comparative study of the RG scale dependence across different approaches. 

\begin{figure}
        \centering
        \includegraphics[width=0.49\linewidth]{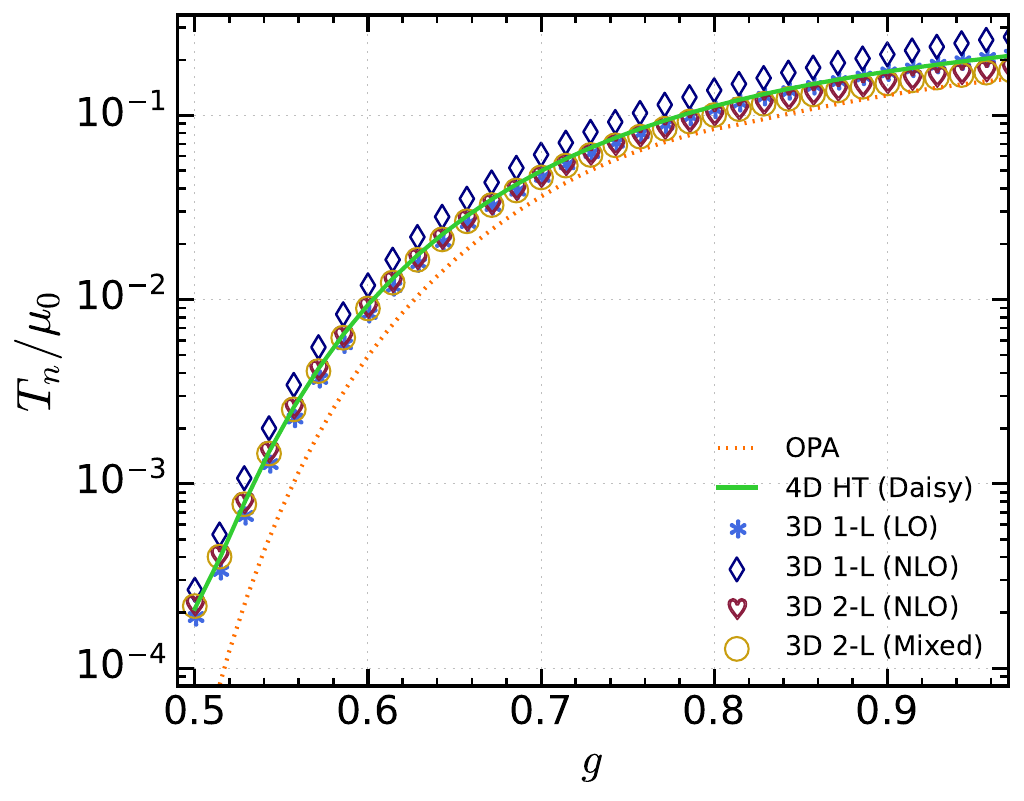}
        \caption{Nucleation temperature $T_n$ as a function of the dark gauge coupling $g$, normalised to the renormalization scale $\mu_0$. Different curves correspond to different computation schemes for the effective potential: 4D HT, OPA, and dimensional reduction.}
        \label{fig:gD_vs_Tn}
    \end{figure}

\Cref{fig:gD_vs_Tn} shows the nucleation temperature $T_n/v$ as a function of the gauge coupling $g$ for different effective potential schemes. The figure indicates that running effects become important at small couplings, $g \lesssim 0.7$, where the OPA approach, which neglects running, deviates significantly from the two-loop result, whereas the HT agrees well with  the two-loop prediction in this regime. For larger couplings, both approximations show comparably sized deviations relative to the two-loop curve.
Having established the impact of the  scheme on the nucleation temperature, we turn to the strength of the transition, quantified by
\begin{equation}
    \alpha = \frac{\Delta V_{\rm eff}^0( g,\mu)}{\rho_R(T_n)} = \frac{V_{\rm eff}^0(0,g,\mu)
    -V_{\rm eff}^0(v,g,\mu) }{\rho_R(T_n)}\, .
\end{equation} 

\begin{figure}
    \centering
    \includegraphics[width=0.49\linewidth]{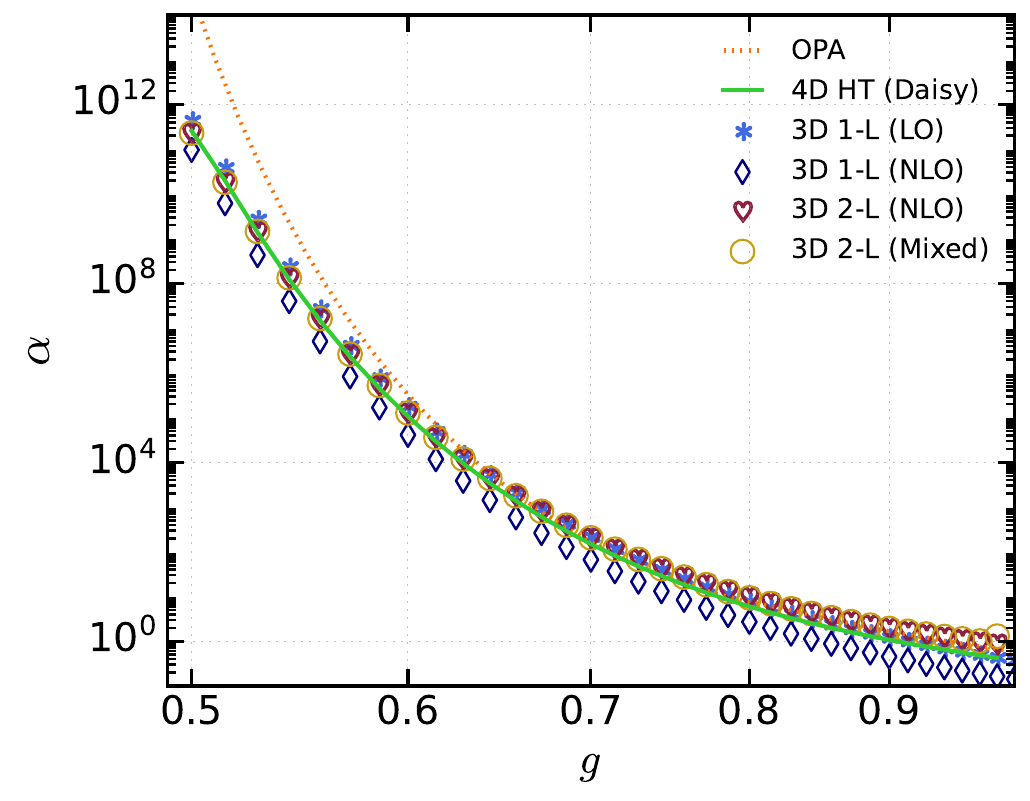}
    \includegraphics[width=0.49\linewidth]{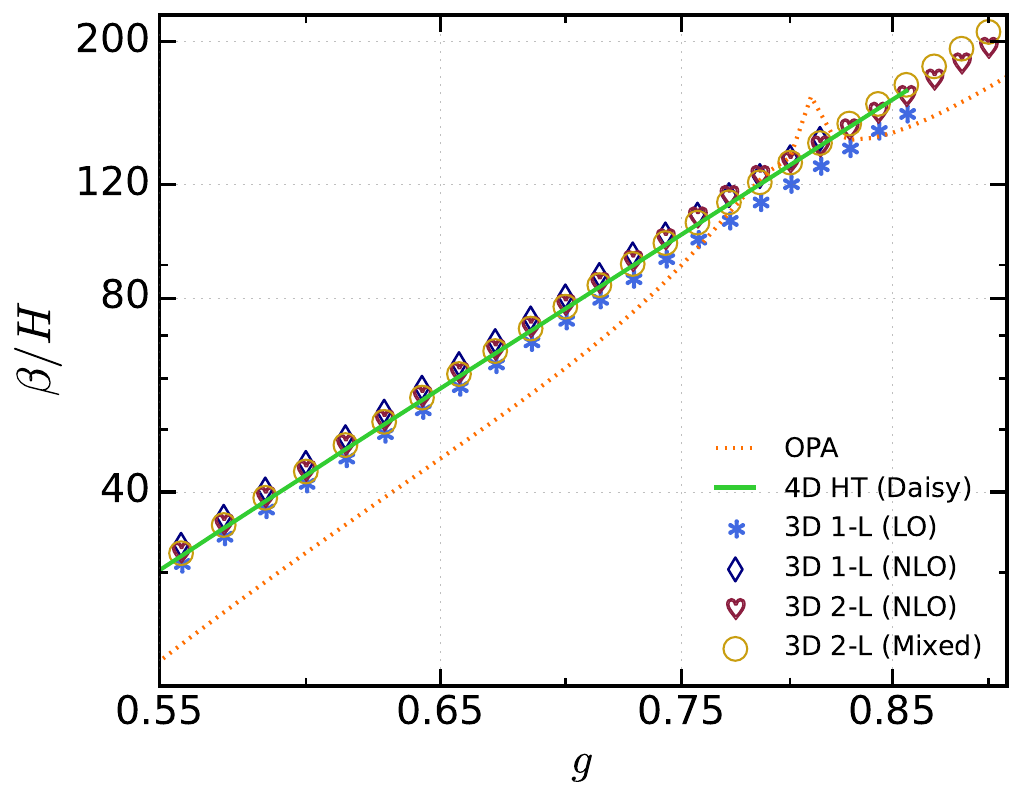}
    \caption{\textbf{Left:} Strength ($\alpha$) and \textbf{Right:} inverse duration  ($\beta/H$) of the phase transition as a function of the dark gauge coupling $g$. Both panels are done for $\mu_0=\SI{1}{\GeV}$ and $\lambda(\mu_0)= 0$. $\alpha$ and $\beta/H$ are computed for different effective potential schemes: 4D HT, OPA and DRalgo.}
    \label{fig:alpha_beta_vs_gD}
\end{figure}

We determine the location of the true minimum from the zero-temperature part of the potential, since the HT and DR methods are only reliable in the vicinity of the thermal barrier. In the supercooling regime the minimum lies far from the barrier, and the HT/DR expansions  may not develop a true minimum, potentially leading to incorrect $\alpha$ values. In such cases, the zero-temperature potential provides a consistent estimate~\cite{Levi:2022bzt}, and we have verified excellent agreement with the minimum obtained from the full 4D potential that does not rely on a high-temperature expansion. The left panel of \cref{fig:alpha_beta_vs_gD} shows $\alpha(g)$; differences between methods, and between including or neglecting running, are minimal, except for OPA and the 3D one-loop (NLO) potential. This agreement is expected because $\alpha$ is extracted from the RG-improved zero-temperature potential evaluated at $\mu = \mu_0$, i.e.\ in the region of the true minimum where thermal corrections and running effects are subdominant, so that deviations between the curves are due to the $T_n$-dependence of the normalization to the radiation energy density $\rho_R(T_n)$.

Finally, we turn to the inverse duration parameter $\beta$ defined by $\beta = {d \log\Gamma}/{d t}$ where $\Gamma$ is the nucleation rate in~\cref{eq:nuc_rate}. 
It is often expressed in dimensionless form as $\beta/H$, where $H$ is the Hubble rate,
    \begin{equation}
    \frac{\beta}{H} =T \frac{d (S_3/T)}{dT} \bigg|_{T_n}.
    \label{eq:beta}
    \end{equation}
This parameter characterises how rapid the PT completes after the onset of nucleation, with larger values of $\beta/H$ corresponding to faster dynamics.  
The right panel of \cref{fig:alpha_beta_vs_gD} shows $\beta/H$ as a function of the gauge coupling $g$. 
In this case, higher-order corrections and running effects have a more sizeable impact. Nevertheless, the 4D HT potential still yields robust results in comparison. By contrast, the OPA curve exhibits a kink, which is unphysical since in this approach the bounce solution is approximated by a  piecewise function~\cite{Levi:2022bzt}, producing this artificial feature, further underlying the limitations of this approach.

\subsection{Gravitational-Wave Spectrum}

\begin{figure}
    \centering
    \includegraphics[width=0.49\linewidth]{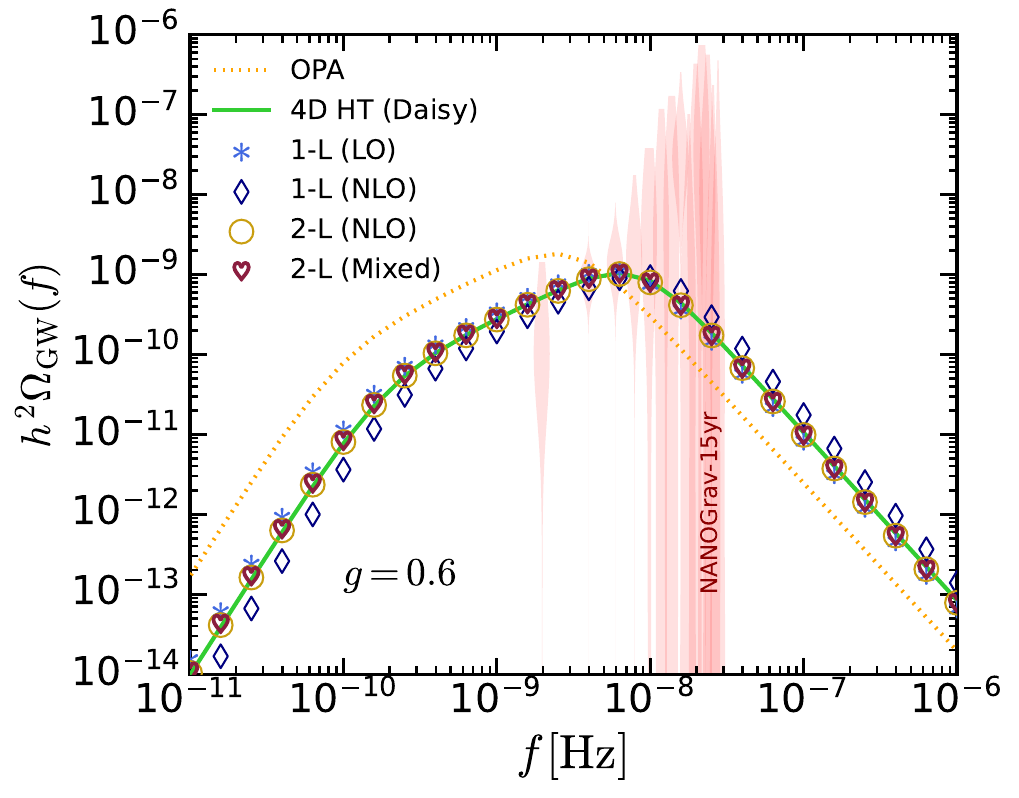}
    \includegraphics[width=0.49\linewidth]{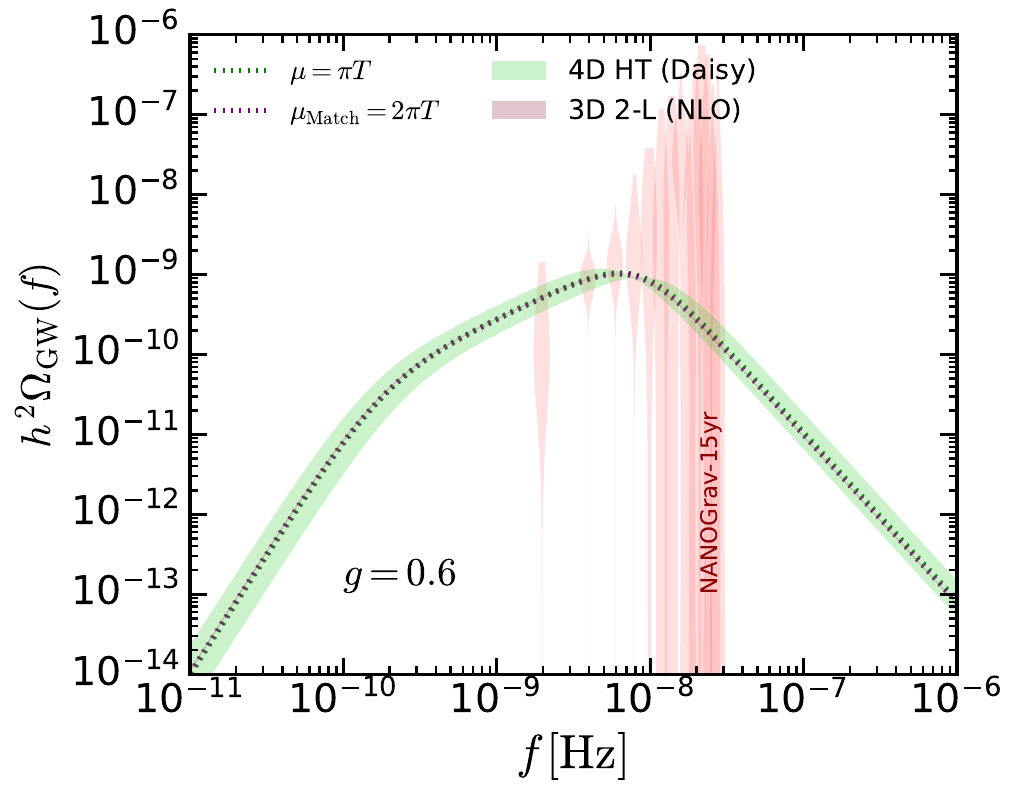}
    \caption{\textbf{Left:} Predicted GW spectrum using the bulk-flow (relativistic) model~\cite{Baldes:2024wuz}, built from phase-transition parameters extracted with different effective-potential schemes for $\mu_0=1$~GeV, and compared to the NANOGrav 15-year posteriors (red violins)~\cite{NANOGrav:2023icp}. The OPA curve visibly departs from both the data and the RGE-improved results, while the 4D HT and 3D DR calculations agree in peak position and height.
    \textbf{Right:} Residual renormalisation-scale dependence propagated to the spectra. As in \cref{fig:Veff_comp} (right), the green (4D HT) and purple (3D two-loop, NLO) bands show the residual $\mu$-dependence, carried into the GW spectra through shifts in $(T_n,\alpha,\beta/H)$. Dotted curves indicate the spectra at the corresponding reference scales.}
    \label{fig:GWs_spect}
\end{figure}

With the FOPT parameters $T_n,\alpha,\beta$ determined%
\footnote{%
    Note that the estimation of these parameters can be further improved, in particular in the supercooled case, by incorporating effects of the expansion of the Universe~\cite{Yamada:2025hfs,Yamada:2025cfr}.
}
and adopting $v_w=1$, we compute the SGWB using the bulk-flow (relativistic) model~\cite{Jinno:2017fby,Konstandin:2017sat,Baldes:2024wuz}.
\Cref{fig:GWs_spect} displays the resulting spectra across the different effective-potential prescriptions, together with the NANOGrav 15-year data~\cite{NANOGrav:2023icp}.

The OPA prediction (orange dashed) visibly departs from the observed signal, highlighting the necessity of incorporating RG-improved dynamics in the determination of the potential. In contrast, both the 4D HT and 3D DR approaches exhibit remarkable consistency in both peak amplitude and peak frequency, reflecting the robustness of the corresponding methods. In particular, the 4D HT potential evaluated at the reference scale $\mu=\pi T$ already provides stable and reliable predictions for the GW spectrum without requiring explicit two-loop corrections. Among the RG-improved schemes, the one-loop (NLO) potential shows the largest deviation, in agreement with the pattern observed in \cref{fig:Veff_comp}, where the absence of higher-order thermal corrections manifests as a residual scale sensitivity.

The right panel further quantifies this residual renormalisation-scale dependence, obtained by propagating the scale variation from the effective potentials into the GW spectra. For the 4D HT computation (green band), we take $\mu=\pi T$ as reference and vary it within $\mu\in[\pi T/4,\,4\pi T]$, while for the 3D two-loop (NLO) result (purple band), we adopt $\mu_{\rm Match}=2\pi T$ and vary it analogously as $\mu_{\rm{Match}}\in[(2\pi T)/4,\,4(2\pi T)]$. In each case, the dotted curve denotes the spectrum evaluated at the reference scale. These bands capture the residual $\mu$-dependence at fixed perturbative order arising from the induced shifts in the phase-transition parameters $(T_n,\alpha,\beta/H)$.  As expected, variations in $T_n$ and $\beta/H$ translate into mild displacements of the peak frequency, whereas the overall amplitude is dominantly controlled by $H/\beta$ and $\alpha$ through the usual efficiency factors.
The reduced width of the 3D two-loop band demonstrates the improved perturbative stability of the dimensionally-reduced framework, consolidating it as a benchmark for precise GW predictions from thermal phase transitions.
%


\section{Conclusions}
\label{sec:conclusions}
In this work we have analysed the theoretical uncertainties associated with different resummation schemes for the finite-temperature effective potential in classically scale-invariant dark photon models undergoing a first-order phase transition. The properties of the transition, and thus the resulting gravitational wave signal, depend sensitively on the treatment of thermal and radiative corrections. We have compared the OPA without running or Daisy corrections, the 4D HT expansion with Daisy resummation, and the 3D DR effective field theory up to two loops. 

Our analysis shows that the inclusion of running couplings is essential for obtaining stable and physically meaningful results. In particular, two-loop DR provides a theoretically robust benchmark, reducing the residual dependence on the renormalisation scale.
We use the two-loop DR result as a controlled benchmark to fix and interpret the renormalisation-scale choice in the 4D one-loop Daisy treatment. With this DR-guided scale setting, the 4D predictions show very good agreement with the two-loop DR results within the expected theory uncertainty. The message is thus not that two-loop DR is unnecessary, but that it provides a principled calibration of the residual $\mu$-dependence in the  4D approach, which is useful for scans and for more involved model extensions.
It should, however, be noted that this conclusion does not necessarily generalise to other models or the non-conformal regime.

A central quantity in our study is the tunnelling action $S_3(T)$, from which we extract the nucleation temperature $T_n$, the strength parameter $\alpha$, and the inverse duration $\beta/H$. We find that the inclusion of running effects can substantially shift $T_n$ and $\beta/H$, while $\alpha$ remains comparatively stable. This highlights the importance of renormalisation-group improvement in connecting microscopic dynamics to macroscopic observables such as the gravitational wave spectrum.

Most notably, the ``one-parameter approximation'' scheme, in which we here neglect running effects, exhibits significant deviations in the predicted FOPT parameters. These discrepancies manifest most clearly in the gravitational wave spectrum: the OPA-based prediction lies further away from the other predictions, emphasising the necessity of incorporating the running for realistic phenomenology. 

Finally, we observe that the next-to-leading-order resummation scheme, while improving certain aspects, introduces additional scale dependence through the matching of the soft scale $\mu_3$ and the couplings at finite temperature. In the absence of higher-order cancellations, this amplifies the sensitivity to $\mu$, making the predictions less stable than in the renormalisation-group improved one-loop 4D HT with Daisy resummation. Overall, our study demonstrates that combining Daisy resummation with RG improvement and an optimised scale choice provides a computationally efficient and theoretically sound framework for predicting gravitational wave signals from supercooled phase transitions, offering a direct and testable link to the PTA observations reported by NANOGrav.

To fully discriminate the phase transition scenario from other primordial 
sources of GWs, complementary probes are required, including laboratory searches 
for the new particles predicted in our model, or other cosmological probes such 
as measurements of CMB spectral distortions~\cite{Ramberg:2022irf}, effective 
number of degrees of freedom, or density fluctuations at small 
scales~\cite{Gouttenoire:2025wxc,Bringmann:2025cht}. Our work will enable a more precise determination 
of the model parameters motivated by the PTA signal, in particular the masses and 
couplings of the Abelian Higgs boson and dark photon in our model~\cite{Madge:2023dxc}. 
We will explore these aspects in future work, where we will also scrutinise the 
uncertainty in predictions of the GW signal from the PT parameters in the strong PT 
regime~\cite{Lewicki:2022pdb,Caprini:2024gyk,Correia:2025qif}.

\section*{Acknowledgements}

MERQ and CPI would like to thank Yann Gouttenoire for sharing NANOGrav 15-year data and for many useful discussions. CPI thanks the participants of the workshop \emph{Gravitational Wave Probes of Physics Beyond the Standard Model 4} (Warsaw) for insightful discussions on the importance of RGE effects, and also thanks Carlo Tasillo, Thomas Konstandin, Maciej Kierkla, Nicklas Ramberg, Philipp Schicho, Daniel Schmitt and Oliver Gould for useful insights.
MERQ gratefully acknowledges M. E. Tejeda-Yeomans for her hospitality at the Universidad de Colima during the final stages of this work.
EM is supported by the Spanish Research Agency (Agencia
Estatal de Investigación) through the grant IFT Centro de Excelencia Severo Ochoa No CEX2020-001007-S
as well as the grants CNS2023-14453 and PID2022-137127NB-I00 funded by MCIN/AEI/10.13039/501100011033, the European Union NextGenerationEU/PRTR and ESF+.
CPI and MERQ are supported by the Cluster of Excellence ``Precision Physics, Fundamental Interactions, and Structure of Matter"  (PRISMA$^+$, EXC 2118/1) funded by the Deutsche Forschungsgemeinschaft (DFG, German Research Foundation) within the German Excellence Strategy (Project No. 390831469).
\appendix
\section{Details of the Effective Potential}
\label{app:Veff_details}

For completeness we collect here the technical ingredients used in the construction of the effective potential. 
\subsection{ Daisy (ring) resummation: 4D potential}
Expanding the complex scalar field of \cref{eq:Lag} as 
$\Phi = (\phi+i\chi)/\sqrt{2}$, we restrict to the real direction $\phi$,
which is the relevant background field for the effective potential. In the 4D approach this can be written schematically as
\begin{align}
    V_{\rm eff}(\phi,T;\mu) = V_{\rm tree}(\phi)
    + V_{\rm CW}(\phi;\mu) 
    + V_T(\phi,T) 
   +  V_{\rm{Daisy}}(\phi,T)\,, 
   \label{eq:Veff}
\end{align}
where  $V_{\rm CW}(\phi;\mu) $ denotes the one-loop quantum corrections,  and $V_T(\phi,T)$ the  thermal contributions. The Daisy term, $V_{\rm{Daisy}}(\phi,T)$, corresponds to the resummation of infrared bosonic modes and is often included as part of the thermal corrections.  Note that $V_{\rm tree}$, $V_T$ and $V_\mathrm{Daisy}$ do not explicitly depend on $\mu$, but implicitly through the  mass $m(\mu)$ and the couplings $g(\mu)$ and $\lambda(\mu)$.

\subsubsection{Zero-Temperature Coleman-Weinberg Potential}
\label{app:Veff_zero}

At zero temperature, the one-loop effective potential in the 
$\overline{\mathrm{MS}}$ scheme takes the form
\begin{equation}
    V_{\rm 1-loop}(\phi) = V_{\rm tree}(\phi)+ V_{\rm CW}(\phi;\mu). 
    \label{eq:1loop_Veff}
\end{equation}
The tree-level scalar potential reads
\begin{align}
    V_{\rm tree}(\phi) = -\frac{m^2}{2}\,\phi^2 
    + \frac{\lambda}{4}\,\phi^4 \,.
\end{align}

and the Coleman-Weinberg one-loop correction in Landau gauge ($\xi=0$) reads,
\begin{align}\begin{aligned}
    V_{\mathrm{\rm CW}}(\phi;\mu) =
     \frac{m_\phi(\phi)^4}{64 \pi^2} 
      \left[\log\frac{m_\phi(\phi)^2}{\mu^2} - \frac{3}{2}\right]
    &+ \frac{ m_\chi(\phi)^4}{64 \pi^2} 
      \left[\log\frac{ m_\chi(\phi)^2}{\mu^2} - \frac{3}{2}\right]
    \\&+ \frac{3   m_A(\phi)^4}{64 \pi^2} \left[\log\frac{  m_A(\phi)^2}{\mu^2} - \frac{5}{6}\right].
      \label{eq:CW_potential}
\end{aligned}\end{align}
The field-dependent masses are 
\begin{align}\begin{aligned}
    m_\phi^2 &= 3 \lambda \phi^2 - m^2 \,,\\
    m_\chi^2 &= \lambda \phi^2 - m^2 \,,\\
    m_A^2 &= g^2 \phi^2. 
    \label{eq:masses}
\end{aligned}\end{align}
These quantities determine the structure of the quantum corrections  entering the Coleman-Weinberg potential. Furthermore, we take the model to be in the scale invariant limit, where we set $m^2(\mu_0)=0$ and take $\lambda(\mu_0)\rightarrow 0$. 
\subsubsection{Thermal Functions and Mass Corrections}

The one-loop thermal potential reads
\begin{equation}
\begin{split}
    V_T(\phi, T) 
    & = \frac{T^4}{2\pi^2}\,J_b(x)
    \,,
\end{split}
\label{eq::V_T}
\end{equation}
where $x= \frac{m_i^2(\phi)}{T^2}$ and the thermal contribution is encoded in the bosonic function
\begin{equation}
      J_b(x)= \int^\infty_0 dk \, k^2 \log\left(1-e^{-\sqrt{k^2+x^2}}\right),
\label{eq:JbInt}
\end{equation}
which admits both numerical evaluation and the high-temperature expansion  \cite{Gouttenoire:2022gwi},

\begin{equation}  
J_b(x) = -\frac{\pi^4}{45}
        + \frac{\pi^2}{12}x
        - \frac{\pi}{6}x^{3/2}
        - \frac{1}{32}x^2\log\!\left(\frac{x}{a_b}\right)
        + \mathcal{O}(x^3 \log x)\,,
        \label{eq:HT_exp}
\end{equation}
where $a_b = 16 \pi^2 \exp(3/2 - 2 \gamma_E)$ with $\gamma_E$ the Euler-Mascheroni constant.

Finally, the thermal mass corrections entering  the Daisy potential, \cref{eq:VDaisy}, are
    \begin{align}
        \Pi_{\phi,\chi} &= \frac{\lambda T^2}{3} + \frac{g^2 T^2}{4}\,, &
        \Pi_{A_L} &= \frac{g^2 T^2}{3}\,.
        \label{eq:thermalmasses}
    \end{align}  

\subsubsection{One-parameter approximation}
\label{app:OPA}

The one-parameter approximation~(OPA) uses that the bounce equation is invariant under reparametrizations of the form $r \to L \rho$ and $\phi \to \xi \varphi$, where $r$ is the radial coordinate, for which the three-dimensional Euclidean action becomes
\begin{equation}
    {S_3} = {4 \pi} \int \!dr\, r^2 \left[ \frac{1}{2} (\partial_r\phi)^2 + V(\phi) \right] = 4 \pi\,\xi^2 L \int \!d\rho\, \rho^2 \left[ \frac{1}{2} (\partial_\rho\varphi)^2 + U(\varphi) \right],
\end{equation}
with $U(\varphi) = L^2 V(\xi \varphi) / \xi^2$.
Using these rescalings, quartic polynomial potentials can be rewritten to depend only on a single parameter $\kappa$,
\begin{equation}
    \label{eq:quartic_potential}
    V(\phi) = \frac{m_T^2}{2}\,\phi^2 - \frac{\delta_T}{3}\,\phi^3 + \frac{\lambda_T}{4}\,\phi^4
    \qquad\longrightarrow\qquad
    U(\varphi) = \frac{1}{2}\,\varphi^2 - \frac{1}{3}\,\varphi^3 + \frac{\kappa_T}{4}\,\varphi^4\,,
\end{equation}
with $\kappa_T = {\lambda_T m_T^2}/{\delta_T^2}$, where we set $\xi = {m_T^2}/{\delta_T}$ and $L = 1/m_T$.\footnote{%
We here assume that $m_T^2>0$, which, due to the smallness of $\lambda(\mu)$, holds in the temperature range considered here.
}
Ref.~\cite{Levi:2022bzt} then provides fits to the rescaled tunneling action $S_3/(\xi^2 L)$ as a function of $\kappa_T$.

In the high-temperature limit, and neglecting Daisy contributions, the scale-invariant CW model with vanishing tachyonic mass parameter $m^2=0$, can be written in the form \cref{eq:quartic_potential} using
\begin{equation}\begin{gathered}
    m_T^2 = \frac{3\,g^2 + 4\,\lambda}{12}\,T^2\,, 
    \qquad
    \delta_T = \frac{3\,g^3 + (1+3\sqrt{3})\,\mathrm{Re}\,\lambda^\frac{3}{2}}{4 \pi}\, T\,, 
    \\
    \lambda_T = \lambda + \frac{3\,g^4}{8 \pi^2}\,\log\frac{T}{M_{g^2}} + \frac{5\,\lambda^2}{4 \pi^2}\,\log\frac{T}{M_\lambda}\,,
\end{gathered}\end{equation}
where $M_{a} = \frac{\mu}{4 \pi}\,\exp(n_a-\gamma_E)$, $a\in\{g^2,\lambda\}$, with $n_{g^2}=1/3$ and $n_\lambda=2/3$. In the limit $\lambda\to 0$, this reproduces the corresponding expression in Ref.~\cite{Levi:2022bzt}.
If the renormalisation scale is set independent of the field value, it is straightforward to incorporate the RG running in the OPA, replacing the couplings by the corresponding running couplings.

The left panel of \cref{fig:beta_Full_HT_OPA} compares the tunneling actions in the OPA and the 4D HT, with and without the inclusion of RG running (for $g(\mu_0)=0.5$). 
While the results in the two approaches differ visibly if running effects are neglected, in particular at high temperatures (close to the critical temperature), the agreement significantly improves once running is taken into account, using the $\mu=\pi T$.
The remaining minor differences between the OPA and 4D HT curves is primarily due to the inclusion of Daisy corrections in the latter.  


\subsubsection{4D potential without running}
    \begin{figure}
        \centering
        \includegraphics[width=0.49\linewidth]{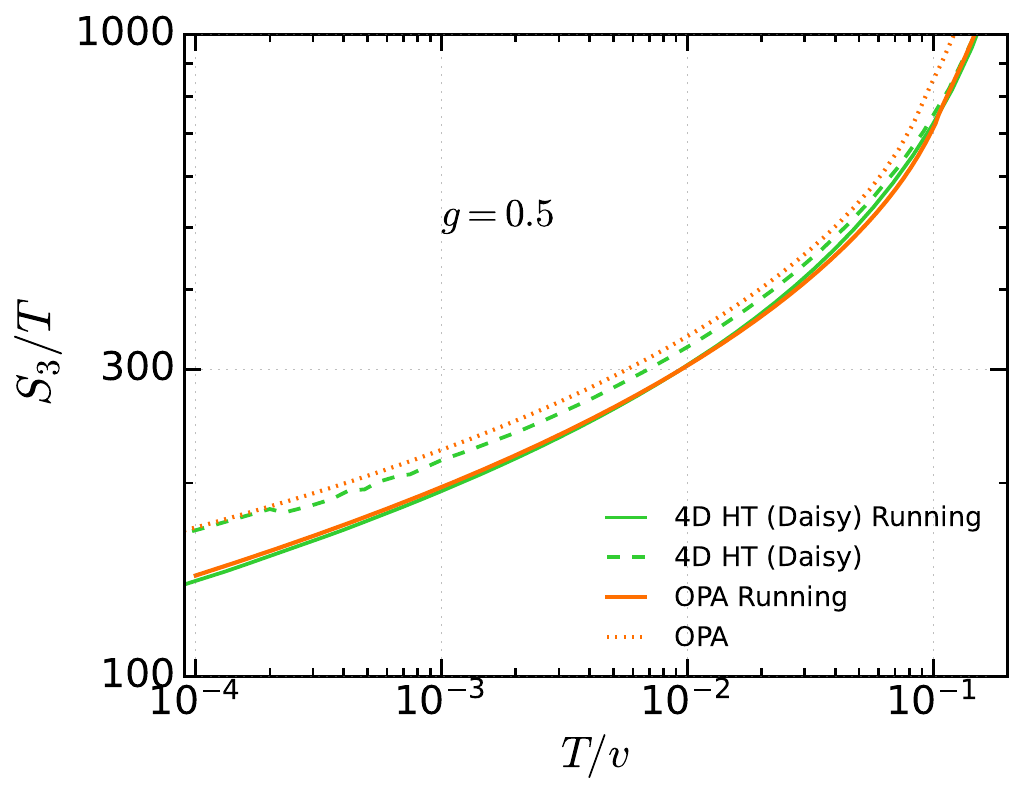}
        \includegraphics[width=0.48\linewidth]{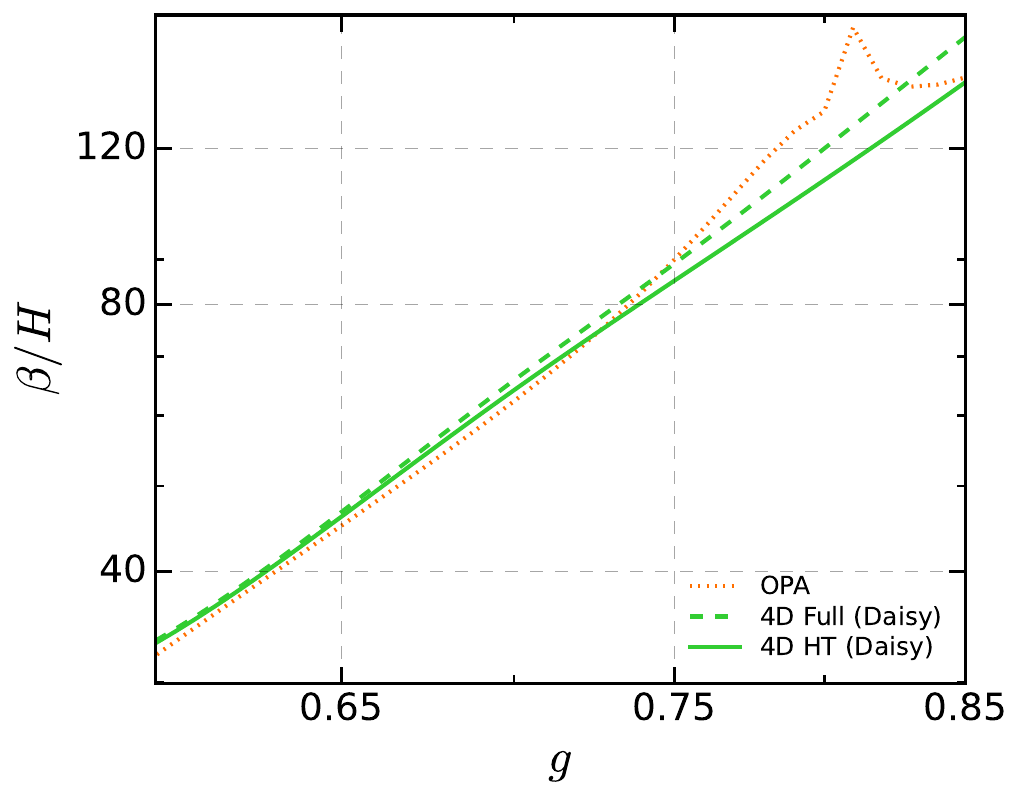}
        
        \caption{%
            \textbf{Left:} Comparison of the OPA (orange) and the 4D HT (green) at a gauge coupling of $g=0.5$ with and without the inclusion of RG running (with $\mu=\pi T$). The two approaches agree well when including the running. The remaining minor differences are primarily due to the inclusion of Daisy corrections in the 4D~HT.
            \textbf{Right:} Duration of the phase transition ($\beta/H$) as a function of the gauge coupling $g$, computed without RGE running. Agreement between 4D Full and 4D~HT improves toward smaller $g$. The OPA deviates at low $g$ because neglecting Daisy resummation overestimates the thermal barrier, lengthening the transition timescale (since $\beta$ is the inverse time). Around $g \simeq 0.8$ the OPA exhibits an unphysical peak, which originates from how the bounce action is defined in that approximation.
        }
        \label{fig:beta_Full_HT_OPA}
    \end{figure}    
A simplified treatment of the 4D potentials fixes the renormalisation scale to a reference value $\mu_0$ and keeps all parameters at their tree-level values (see the last row of \cref{tab:mu_scales}), thus neglecting RG running. This setup coincides with what is implicitly used in the OPA throughout most of this paper.

For supercooled transitions, extensive pre-studies showed that 4D Full and 4D HT yield nearly identical results for all transition parameters and for the resulting GW spectra. The agreement even improves at lower transition temperatures: when the thermal barrier is small and the bounce occurs close to the origin, the high-temperature expansion remains valid; see, e.g., \cref{fig:beta_Full_HT_OPA}.

Because 4D HT reproduces 4D Full while being computationally cheaper, we adopt it as our reference 4D scheme with Daisy resummation in the main analysis. In contrast, the OPA performs visibly worse. Its missing Daisy corrections are the dominant source of deviation, and its  piecewise fit to the bounce introduces artifacts --- most notably the unphysical peak in $\beta/H$ around $g \simeq 0.8$ in \cref{fig:beta_Full_HT_OPA}.

\subsection{Dimensional Reduction: 3D EFT Parameters}
\label{app:VeffDR}
Dimensional reduction integrates out the non-zero Matsubara modes, leading to a three-dimensional EFT for the soft modes.  
The corresponding Lagrangian is
\begin{align}\begin{aligned}
    \mathcal{L}_\soft &= \frac{1}{4} {{F}^\soft_{ij}}^2 
    + \frac{1}{2} \left(\partial_i A_0^\soft\right)^2 
    + \frac{1}{2} \left|\left(\partial_i - i g_\soft A^\soft_i\right)\Phi^\soft\right|^2 
    - m_\soft^2 \left|\Phi^\soft\right|^2 \\
    &\phantom{=}\ 
    + \frac{1}{2} \muD^2 {A_0^\soft}^2 
    + \lambda_\soft \left|\Phi^\soft\right|^4 
    + \frac{\lamVLL}{4!} {A_0^\soft}^4
    + \frac{\lamVL}{2} {A_0^\soft}^2 \left|\Phi^\soft\right|^2 \,.
    \label{eq:L_DR}
\end{aligned}\end{align}
We follow the conventions of \texttt{DRalgo}~\cite{Ekstedt:2022bff}, which provides the matching relations at one loop for couplings and up to two loops for mass parameters.  At one-loop, the parameters of the 3D EFT are
\begin{align}\begin{aligned}
    g_\soft^2 &= g^2 T - \frac{g^4 L_b T}{48 \pi^2} \,, \qquad&
    \lambda_\soft &= \lambda T + \frac{T}{16 \pi^2} \left[ g^4 \left( 2 - 3 L_b \right) + 6 g^2 \lambda L_b  - 10 \lambda^2 L_b \right]\,, \\
    \lambda_{A} &= 0 + \frac{g^4 T}{\pi^2}\,,&
    \lambda_{A\Phi} &= 2 g^2 T + \frac{g^2 T}{24 \pi^2} \left[ 24 \lambda - g^2 \left(L_b-4\right) \right]\,,\label{eq:3dSoftCouplings}
\end{aligned}\end{align}
where
\begin{equation}
    L_b = \log\frac{\mu^2 e^{2 \gamma_E}}{16 \pi^2 T^2}
\end{equation}

The scalar mass parameter at two-loop order written in terms of the 3D EFT couplings is
\begin{align}\begin{aligned}
    m_\soft^2 =
        m^2 - \frac{T^2 (3 g^2 + 4 \lambda)}{12} 
        &+ \frac{L_b ( 3 g^2 - 4 \lambda ) m^2}{16 \pi^2} 
        + \frac{g^4 (8 + 216 c_+ + 39 L_b) T^2}{576 \pi^2} \\[2mm]
        &+ \frac{\lambda^2 (12 c_+ + 5 L_b) T^2}{24 \pi^2} 
        - \frac{g^2 \lambda (1 + 12 c_+ + 3 L_b) T^2}{24 \pi^2} \\[2mm]
        &- \frac{8 g_\soft^4 - 16 g_\soft^2 \lambda_\soft + 16 \lambda_\soft^2 + \lamVL^2}{32 \pi^2} 
        \log\frac{\mu_3}{\mu} \,.
         \label{eq:msq3d2loop_3d}
\end{aligned}\end{align}
Here $\mu_3$ is the 3D renormalisation scale, $c_+ = \tfrac{1}{2}\left(\gamma_E - L_b - 12 \log A \right)$ with $A$ the Glaisher-Kinkelin constant.
For practical applications, it is often convenient to re-express the above relation directly in terms of the 4D couplings. In this case we obtain
\begin{align}\begin{aligned}
    m_\soft^2 =
        m^2 - \frac{T^2 (3 g^2 + 4 \lambda)}{12} 
        +& \frac{L_b ( 3 g^2 - 4 \lambda ) m^2 }{16 \pi^2}
        + \frac{L_b ( 13 g^4 - 24 g^2 \lambda + 40 \lambda^2 ) T^2}{192 \pi^2} \\[2mm]
        &\phantom{=} + \frac{c_3 (3 g^4 - 4 g^2 \lambda + 4 \lambda^2) T^2}{8 \pi^2}
        + \frac{(g^4 - 3 g^2 \lambda) T^2}{72 \pi^2} \,,
        \label{eq:msq3d2loop_4d}
\end{aligned}\end{align}
with $c_3 = c_+ - \log(\mu_3/\mu)$. 
Note that in \cref{eq:msq3d2loop_3d} we replaced the 3D couplings inside the logarithm by their tree-level values 
($g^2_\soft \to g^2 T$, $\lambda_\soft \to \lambda T$, $\lamVL \to 2 g^2 T$) 
for consistency, assuming $g^2\sim\lambda\sim m^2/T^2$ in the power counting.

The temporal component of the gauge boson acquires an effective thermal mass (Debye mass). At two-loop order it reads
\begin{align}
    \label{eq:DebyeMass}
    \muD^2 = \frac{g^2 T^2}{3} + \frac{g^2}{144 \pi^2} \left[ 12 \lambda T^2 + g^2 (7-L_b) T^2 - 36 m^2 \right] \,.
\end{align}

\texttt{DRalgo} provides the effective potential up to two loops, with the tree-level, one-loop and two-loop contributions starting at order~$g^2$, $g^3$, and $g^4$, respectively (taking $\lambda \sim m^2/T^2 \sim \mathcal{O}(g^2)$).
We then have two options to evaluate the two-loop effective potential in the soft theory:
\begin{enumerate}
    \item \textbf{2-Loop (NLO):} calculate the 3D couplings and masses at NLO (one-loop for the couplings and two-loop for the masses), i.e.\ Eqs.~\eqref{eq:3dSoftCouplings} and \eqref{eq:msq3d2loop_3d} (replacing the 3D couplings in the latter by the full one-loop couplings Eq.~\eqref{eq:3dSoftCouplings}) and Eq.~\eqref{eq:DebyeMass}, and plug these into the tree-level + one-loop + two-loop soft effective potential. 
    \item \textbf{2-Loop (Mixed):} replace the masses and couplings in the tree-level soft potential by one-loop couplings and two-loop masses in Eqs.~\eqref{eq:3dSoftCouplings}, \eqref{eq:msq3d2loop_4d} and \eqref{eq:DebyeMass}, but evaluate the one- and two-loop parts of the potential only with the leading-order expressions.
\end{enumerate}

For the full NLO scheme, the scalar mass and Debye mass at leading order are
\begin{align}
    m_{3D}^2 &= m^2 - \frac{T^2 (3 g^2 + 4 \lambda)}{12}, 
    & \quad \mu_D^2 &= \frac{g^2 T^2}{3}\,,
\end{align}
while the one-loop couplings read
\begin{align}
    g_{3D}^2 &= g^2 T - \frac{g^4 L_b T}{48 \pi^2}, \\
    \lambda_{S3D} &= \frac{T \big( g^4 (2 - 3 L_b) + 6 g^2 L_b \lambda + 2 \lambda ( 8 \pi^2 - 5 L_b \lambda ) \big)}{16 \pi^2}, \\
    \lambda_{A\Phi} &= \frac{g^2 T \big( -g^2 (-4 + L_b) + 24 (2 \pi^2 + \lambda) \big)}{24 \pi^2}.
\end{align}
Note that the Debye mass coincides with the leading 4D thermal correction to the longitudinal boson, $\Pi_{A_L}$, which appears in the Daisy resummation of \cref{eq:VDaisy}. 

For the truncated scheme, in order to match the 4D high-temperature potential, we use the same masses as above but truncate the couplings as
\begin{align}
    g_{3D}^2 &= g^2 T, \qquad \lambda_{A\Phi} = 2 g^2 T,
\end{align}
while for $\lambda_{S3D}$ we keep
\begin{align}
    \lambda_{S3D} = 
    \begin{cases}
       \dfrac{T \left( g^4 ( 2 - 3 L_b ) + 2 \lambda ( 8 \pi^2 - 5 L_b \lambda ) \right)}{16 \pi^2}, & \text{tree level}, \\[1.2ex]
       T \lambda, & \text{LO potential}.
    \end{cases}
\end{align}

\section{4D Running Couplings and Anomalous Dimensions}
\label{app:Beta_functions}

The one-loop running of the couplings is governed by
\begin{align}
    \beta_{g^2} &= \frac{g^4}{24 \pi^2}, &
    \beta_{\lambda} &= \frac{10 \lambda^2 - 6 \lambda g^2 + 3 g^4}{8 \pi^2}, &
    \beta_{m^2} &= m^2 \frac{4 \lambda - 3 g^2}{8 \pi^2}, 
    \label{eq:betas}
\end{align}
together with the anomalous dimensions
\begin{align}
    \gamma_\Phi = \frac{3 g^2}{16 \pi^2}, 
    \qquad
    \gamma_A = -\frac{g^2}{48 \pi^2}.
    \label{eq:gammas}
\end{align}

These relations imply, for example, the running of the gauge coupling,
\begin{align}\label{eq:fine-structure}
    \alpha(\mu) = \frac{g^2(\mu)}{4 \pi} 
    = \frac{\alpha_0}{1 - \tfrac{\alpha_0}{6 \pi}\log(\mu/\mu_0)} \,,
\end{align}
and the corresponding wave-function renormalisation of the scalar,
\begin{align}
    Z_\Phi(\mu) = Z_0 \left(\frac{\alpha_0}{\alpha(\mu)}\right)^{9}.
\end{align}

For the RG improvement of the effective potential we adopt the usual choice of renormalisation scale $\mu = \pi T$, which minimises large logarithms in the relevant regimes.

\bibliographystyle{JHEP}
\bibliography{lib_jhep}

@article{Balan:2025uke,
    author = "Balan, Sowmiya and Bringmann, Torsten and Kahlhoefer, Felix and Matuszak, Jonas and Tasillo, Carlo",
    title = "{Sub-GeV dark matter and nano-Hertz gravitational waves from a classically conformal dark sector}",
    eprint = "2502.19478",
    archivePrefix = "arXiv",
    primaryClass = "hep-ph",
    doi = "10.1088/1475-7516/2025/08/062",
    journal = "J. Cosmol. Astropart. Phys.",
    volume = "08",
    pages = "062",
    year = "2025",
    number = "2025"
}

@article{Gouttenoire:2025wxc,
    author = "Gouttenoire, Yann",
    title = "{WIMPs and new physics interpretations of the PTA signal are incompatible}",
    eprint = "2503.03857",
    archivePrefix = "arXiv",
    primaryClass = "hep-ph",
    month = "3",
    year = "2025"
}

@article{Ellis:2023oxs,
    author = {Ellis, John and Fairbairn, Malcolm and Franciolini, Gabriele and H{\"u}tsi, Gert and Iovino, Antonio and Lewicki, Marek and Raidal, Martti and Urrutia, Juan and Vaskonen, Ville and Veerm{\"a}e, Hardi},
    title = "{What is the source of the PTA GW signal?}",
    eprint = "2308.08546",
    archivePrefix = "arXiv",
    primaryClass = "astro-ph.CO",
    reportNumber = "KCL-PH-TH/2023-43, CERN-TH-2023-153, AION-REPORT/2023-08",
    doi = "10.1103/PhysRevD.109.023522",
    journal = "Phys. Rev. D",
    volume = "109",
    number = "2",
    pages = "023522",
    year = "2024"
}

@article{Madge:2023dxc,
    author = "Madge, Eric and Morgante, Enrico and Puchades-Ib{\'a}{\~n}ez, Cristina and Ramberg, Nicklas and Ratzinger, Wolfram and Schenk, Sebastian and Schwaller, Pedro",
    title = "{Primordial gravitational waves in the nano-Hertz regime and PTA data {\textemdash} towards solving the GW inverse problem}",
    eprint = "2306.14856",
    archivePrefix = "arXiv",
    primaryClass = "hep-ph",
    reportNumber = "MITP-23-029",
    doi = "10.1007/JHEP10(2023)171",
    journal = "J. High Energy Phys.",
    volume = "10",
    pages = "171",
    year = "2023",
    number = "2023"
}

@article{Espinosa:2018hue,
    author = "Espinosa, J. R.",
    title = "{A Fresh Look at the Calculation of Tunneling Actions}",
    eprint = "1805.03680",
    archivePrefix = "arXiv",
    primaryClass = "hep-th",
    doi = "10.1088/1475-7516/2018/07/036",
    journal = "J. Cosmol. Astropart. Phys.",
    volume = "07",
    pages = "036",
    year = "2018",
    number = "2018"
}

@article{Marzo:2018nov,
    author = "Marzo, Carlo and Marzola, Luca and Vaskonen, Ville",
    title = "{Phase transition and vacuum stability in the classically conformal B\textendash{}L model}",
    eprint = "1811.11169",
    archivePrefix = "arXiv",
    primaryClass = "hep-ph",
    reportNumber = "KCL-PH-TH/2018-68",
    doi = "10.1140/epjc/s10052-019-7076-x",
    journal = "Eur. Phys. J. C",
    volume = "79",
    number = "7",
    pages = "601",
    year = "2019"
}

@book{Gouttenoire:2022gwi,
    author = "Gouttenoire, Yann",
    title = "{Beyond the Standard Model Cocktail}",
    eprint = "2207.01633",
    archivePrefix = "arXiv",
    primaryClass = "hep-ph",
    doi = "10.1007/978-3-031-11862-3",
    isbn = "978-3-031-11862-3, 978-3-031-11861-6",
    publisher = "Springer",
    address = "Cham",
    series = "Springer Theses",
    year = "2022"
}

@article{Levi:2022bzt,
    author = "Levi, Noam and Opferkuch, Toby and Redigolo, Diego",
    title = "{The supercooling window at weak and strong coupling}",
    eprint = "2212.08085",
    archivePrefix = "arXiv",
    primaryClass = "hep-ph",
    doi = "10.1007/JHEP02(2023)125",
    journal = "J. High Energy Phys.",
    volume = "02",
    pages = "125",
    year = "2023",
    number = "2023"
}

@article{Coleman:1973jx,
    author = "Coleman, Sidney R. and Weinberg, Erick J.",
    title = "{Radiative Corrections as the Origin of Spontaneous Symmetry Breaking}",
    doi = "10.1103/PhysRevD.7.1888",
    journal = "Phys. Rev. D",
    volume = "7",
    pages = "1888--1910",
    year = "1973"
}

@article{Ekstedt:2022bff,
    author = "Ekstedt, Andreas and Schicho, Philipp and Tenkanen, Tuomas V. I.",
    title = "{DRalgo: A package for effective field theory approach for thermal phase transitions}",
    eprint = "2205.08815",
    archivePrefix = "arXiv",
    primaryClass = "hep-ph",
    reportNumber = "HIP-2022-11/TH, NORDITA 2022-030",
    doi = "10.1016/j.cpc.2023.108725",
    journal = "Comput. Phys. Commun.",
    volume = "288",
    pages = "108725",
    year = "2023"
}

@article{Matsubara:1955ws,
    author = "Matsubara, Takeo",
    title = "{A New approach to quantum statistical mechanics}",
    doi = "10.1143/PTP.14.351",
    journal = "Prog. Theor. Phys.",
    volume = "14",
    pages = "351--378",
    year = "1955"
}

@article{Kierkla:2023von,
    author = "Kierkla, Maciej and Swiezewska, Bogumila and Tenkanen, Tuomas V. I. and van de Vis, Jorinde",
    title = "{Gravitational waves from supercooled phase transitions: dimensional transmutation meets dimensional reduction}",
    eprint = "2312.12413",
    archivePrefix = "arXiv",
    primaryClass = "hep-ph",
    doi = "10.1007/JHEP02(2024)234",
    journal = "J. High Energy Phys.",
    volume = "02",
    pages = "234",
    year = "2024",
    number = "2024"
}

@article{Gould:2021oba,
    author = "Gould, Oliver and Tenkanen, Tuomas V. I.",
    title = "{On the perturbative expansion at high temperature and implications for cosmological phase transitions}",
    eprint = "2104.04399",
    archivePrefix = "arXiv",
    primaryClass = "hep-ph",
    reportNumber = "NORDITA 2021-010",
    doi = "10.1007/JHEP06(2021)069",
    journal = "J. High Energy Phys.",
    volume = "06",
    pages = "069",
    year = "2021",
    number = "2021"
}

@article{Lewicki:2022pdb,
    author = "Lewicki, Marek and Vaskonen, Ville",
    title = "{Gravitational waves from bubble collisions and fluid motion in strongly supercooled phase transitions}",
    eprint = "2208.11697",
    archivePrefix = "arXiv",
    primaryClass = "astro-ph.CO",
    doi = "10.1140/epjc/s10052-023-11241-3",
    journal = "Eur. Phys. J. C",
    volume = "83",
    number = "2",
    pages = "109",
    year = "2023"
}

@article{Baldes:2024wuz,
    author = "Baldes, Iason and Dichtl, Maximilian and Gouttenoire, Yann and Sala, Filippo",
    title = "{Particle shells from relativistic bubble walls}",
    eprint = "2403.05615",
    archivePrefix = "arXiv",
    primaryClass = "hep-ph",
    doi = "10.1007/JHEP07(2024)231",
    journal = "J. High Energy Phys.",
    volume = "07",
    pages = "231",
    year = "2024",
    number = "2024"
}

@article{Arnold:1992rz,
    author = "Arnold, Peter Brockway and Espinosa, Olivier",
    title = "{The Effective potential and first order phase transitions: Beyond leading-order}",
    eprint = "hep-ph/9212235",
    archivePrefix = "arXiv",
    reportNumber = "UW-PT-92-18, USM-TH-60",
    doi = "10.1103/PhysRevD.47.3546",
    journal = "Phys. Rev. D",
    volume = "47",
    pages = "3546",
    year = "1993",
    note = "[Erratum: Phys.Rev.D 50, 6662 (1994)]"
}

@article{NANOGrav:2021flc,
    author = "Arzoumanian, Zaven and others",
    collaboration = "NANOGrav",
    title = "{Searching for Gravitational Waves from Cosmological Phase Transitions with the NANOGrav 12.5-Year Dataset}",
    eprint = "2104.13930",
    archivePrefix = "arXiv",
    primaryClass = "astro-ph.CO",
    doi = "10.1103/PhysRevLett.127.251302",
    journal = "Phys. Rev. Lett.",
    volume = "127",
    number = "25",
    pages = "251302",
    year = "2021"
}

@article{NANOGrav:2023icp,
    author = "Johnson, Aaron D. and others",
    collaboration = "NANOGrav",
    title = "{NANOGrav 15-year gravitational-wave background methods}",
    eprint = "2306.16223",
    archivePrefix = "arXiv",
    primaryClass = "astro-ph.HE",
    doi = "10.1103/PhysRevD.109.103012",
    journal = "Phys. Rev. D",
    volume = "109",
    number = "10",
    pages = "103012",
    year = "2024"
}

@article{NANOGrav:2023hvm,
    author = "Afzal, Adeela and others",
    collaboration = "NANOGrav",
    title = "{The NANOGrav 15 yr Data Set: Search for Signals from New Physics}",
    eprint = "2306.16219",
    archivePrefix = "arXiv",
    primaryClass = "astro-ph.HE",
    reportNumber = "FERMILAB-PUB-23-589-T",
    doi = "10.3847/2041-8213/acdc91",
    journal = "Astrophys. J. Lett.",
    volume = "951",
    number = "1",
    pages = "L11",
    year = "2023",
    note = "[Erratum: Astrophys.J.Lett. 971, L27 (2024), Erratum: Astrophys.J. 971, L27 (2024)]"
}

@article{LIGOScientific:2016aoc,
    author = "Abbott, B. P. and others",
    collaboration = "LIGO Scientific, Virgo",
    title = "{Observation of Gravitational Waves from a Binary Black Hole Merger}",
    eprint = "1602.03837",
    archivePrefix = "arXiv",
    primaryClass = "gr-qc",
    reportNumber = "LIGO-P150914",
    doi = "10.1103/PhysRevLett.116.061102",
    journal = "Phys. Rev. Lett.",
    volume = "116",
    number = "6",
    pages = "061102",
    year = "2016"
}

@article{Kramer:2013kea,
    author = "Kramer, Michael and Champion, David J.",
    title = "{The European Pulsar Timing Array and the Large European Array for Pulsars}",
    doi = "10.1088/0264-9381/30/22/224009",
    journal = "Class. Quant. Grav.",
    volume = "30",
    pages = "224009",
    year = "2013"
}

@article{Desvignes:2016yex,
    author = "Desvignes, G. and others",
    title = "{High-precision timing of 42 millisecond pulsars with the European Pulsar Timing Array}",
    eprint = "1602.08511",
    archivePrefix = "arXiv",
    primaryClass = "astro-ph.HE",
    doi = "10.1093/mnras/stw483",
    journal = "Mon. Not. Roy. Astron. Soc.",
    volume = "458",
    number = "3",
    pages = "3341--3380",
    year = "2016"
}

@article{McLaughlin:2013ira,
    author = "McLaughlin, Maura A.",
    title = "{The North American Nanohertz Observatory for Gravitational Waves}",
    eprint = "1310.0758",
    archivePrefix = "arXiv",
    primaryClass = "astro-ph.IM",
    doi = "10.1088/0264-9381/30/22/224008",
    journal = "Class. Quant. Grav.",
    volume = "30",
    pages = "224008",
    year = "2013"
}

@article{Becsy:2022pnr,
    author = "B\'ecsy, Bence and Cornish, Neil J. and Kelley, Luke Zoltan",
    title = "{Exploring Realistic Nanohertz Gravitational-wave Backgrounds}",
    eprint = "2207.01607",
    archivePrefix = "arXiv",
    primaryClass = "astro-ph.HE",
    doi = "10.3847/1538-4357/aca1b2",
    journal = "Astrophys. J.",
    volume = "941",
    number = "2",
    pages = "119",
    year = "2022"
}

@article{Ellis:2023owy,
    author = {Ellis, John and Fairbairn, Malcolm and H{\"u}tsi, Gert and Raidal, Martti and Urrutia, Juan and Vaskonen, Ville and Veerm{\"a}e, Hardi},
    title = "{Prospects for future binary black hole gravitational wave studies in light of PTA measurements}",
    eprint = "2301.13854",
    archivePrefix = "arXiv",
    primaryClass = "astro-ph.CO",
    reportNumber = "KCL-PH-TH/2023-04, CERN-TH-2023-008, AION-REPORT/2023-1",
    doi = "10.1051/0004-6361/202346268",
    journal = "Astron. Astrophys.",
    volume = "676",
    pages = "A38",
    year = "2023"
}

@article{Nakai:2020oit,
    author = "Nakai, Yuichiro and Suzuki, Motoo and Takahashi, Fuminobu and Yamada, Masaki",
    title = "{Gravitational Waves and Dark Radiation from Dark Phase Transition: Connecting NANOGrav Pulsar Timing Data and Hubble Tension}",
    eprint = "2009.09754",
    archivePrefix = "arXiv",
    primaryClass = "astro-ph.CO",
    reportNumber = "TU-1109; IPMU20-0100",
    doi = "10.1016/j.physletb.2021.136238",
    journal = "Phys. Lett. B",
    volume = "816",
    pages = "136238",
    year = "2021"
}

@article{Ratzinger:2020koh,
    author = "Ratzinger, Wolfram and Schwaller, Pedro",
    title = "{Whispers from the dark side: Confronting light new physics with NANOGrav data}",
    eprint = "2009.11875",
    archivePrefix = "arXiv",
    primaryClass = "astro-ph.CO",
    reportNumber = "MITP/20-056",
    doi = "10.21468/SciPostPhys.10.2.047",
    journal = "SciPost Phys.",
    volume = "10",
    number = "2",
    pages = "047",
    year = "2021"
}

@article{Freese:2022qrl,
    author = "Freese, Katherine and Winkler, Martin Wolfgang",
    title = "{Have pulsar timing arrays detected the hot big bang: Gravitational waves from strong first order phase transitions in the early Universe}",
    eprint = "2208.03330",
    archivePrefix = "arXiv",
    primaryClass = "astro-ph.CO",
    reportNumber = "UTTG 11-2022, NORDITA 2022-056",
    doi = "10.1103/PhysRevD.106.103523",
    journal = "Phys. Rev. D",
    volume = "106",
    number = "10",
    pages = "103523",
    year = "2022"
}

@article{Morgante:2022zvc,
    author = "Morgante, Enrico and Ramberg, Nicklas and Schwaller, Pedro",
    title = "{Gravitational waves from dark SU(3) Yang-Mills theory}",
    eprint = "2210.11821",
    archivePrefix = "arXiv",
    primaryClass = "hep-ph",
    doi = "10.1103/PhysRevD.107.036010",
    journal = "Phys. Rev. D",
    volume = "107",
    number = "3",
    pages = "036010",
    year = "2023"
}

@article{Bringmann:2023opz,
    author = "Bringmann, Torsten and Depta, Paul Frederik and Konstandin, Thomas and Schmidt-Hoberg, Kai and Tasillo, Carlo",
    title = "{Does NANOGrav observe a dark sector phase transition?}",
    eprint = "2306.09411",
    archivePrefix = "arXiv",
    primaryClass = "astro-ph.CO",
    reportNumber = "DESY-23-077",
    doi = "10.1088/1475-7516/2023/11/053",
    journal = "J. Cosmol. Astropart. Phys.",
    volume = "11",
    pages = "053",
    year = "2023",
    number = "2023"
}

@article{Ashoorioon:2022raz,
    author = "Ashoorioon, Amjad and Rezazadeh, Kazem and Rostami, Abasalt",
    title = "{NANOGrav signal from the end of inflation and the LIGO mass and heavier primordial black holes}",
    eprint = "2202.01131",
    archivePrefix = "arXiv",
    primaryClass = "astro-ph.CO",
    reportNumber = "IPM/P-2022/06",
    doi = "10.1016/j.physletb.2022.137542",
    journal = "Phys. Lett. B",
    volume = "835",
    pages = "137542",
    year = "2022"
}

@article{Ferreira:2022zzo,
    author = "Ferreira, Ricardo Z. and Notari, Alessio and Pujolas, Oriol and Rompineve, Fabrizio",
    title = "{Gravitational waves from domain walls in Pulsar Timing Array datasets}",
    eprint = "2204.04228",
    archivePrefix = "arXiv",
    primaryClass = "astro-ph.CO",
    reportNumber = "CERN-TH-2022-214",
    doi = "10.1088/1475-7516/2023/02/001",
    journal = "J. Cosmol. Astropart. Phys.",
    volume = "02",
    pages = "001",
    year = "2023",
    number = "2023"
}

@article{Croon:2020cgk,
    author = "Croon, Djuna and Gould, Oliver and Schicho, Philipp and Tenkanen, Tuomas V. I. and White, Graham",
    title = "{Theoretical uncertainties for cosmological first-order phase transitions}",
    eprint = "2009.10080",
    archivePrefix = "arXiv",
    primaryClass = "hep-ph",
    reportNumber = "HIP-2020-26/TH",
    doi = "10.1007/JHEP04(2021)055",
    journal = "J. High Energy Phys.",
    volume = "04",
    pages = "055",
    year = "2021",
    number = "2021"
}

@article{Kajantie:1996mn,
    author = "Kajantie, K. and Laine, M. and Rummukainen, K. and Shaposhnikov, Mikhail E.",
    title = "{Is there a~ hot electroweak phase transition at $m_H \gtrsim m_W$?}",
    eprint = "hep-ph/9605288",
    archivePrefix = "arXiv",
    reportNumber = "CERN-TH-96-126, HD-THEP-96-15, IUHET-333",
    doi = "10.1103/PhysRevLett.77.2887",
    journal = "Phys. Rev. Lett.",
    volume = "77",
    pages = "2887--2890",
    year = "1996"
}

@article{Aoki:2006we,
    author = "Aoki, Y. and Endrodi, G. and Fodor, Z. and Katz, S. D. and Szabo, K. K.",
    title = "{The Order of the quantum chromodynamics transition predicted by the standard model of particle physics}",
    eprint = "hep-lat/0611014",
    archivePrefix = "arXiv",
    doi = "10.1038/nature05120",
    journal = "Nature",
    volume = "443",
    pages = "675--678",
    year = "2006"
}

@article{Ellis:2019oqb,
    author = "Ellis, John and Lewicki, Marek and No, Jos{\'e} Miguel and Vaskonen, Ville",
    title = "{Gravitational wave energy budget in strongly supercooled phase transitions}",
    eprint = "1903.09642",
    archivePrefix = "arXiv",
    primaryClass = "hep-ph",
    reportNumber = "KCL-PH-TH/2019-32, CERN-TH-2019-032, IFT-UAM/CSIC-19-32",
    doi = "10.1088/1475-7516/2019/06/024",
    journal = "J. Cosmol. Astropart. Phys.",
    volume = "06",
    pages = "024",
    year = "2019",
    number = "2019"
}

@article{Ellis:2020nnr,
    author = "Ellis, John and Lewicki, Marek and Vaskonen, Ville",
    title = "{Updated predictions for gravitational waves produced in a strongly supercooled phase transition}",
    eprint = "2007.15586",
    archivePrefix = "arXiv",
    primaryClass = "astro-ph.CO",
    reportNumber = "KCL-PH-TH/2020-40, CERN-TH-2020-129",
    doi = "10.1088/1475-7516/2020/11/020",
    journal = "J. Cosmol. Astropart. Phys.",
    volume = "11",
    pages = "020",
    year = "2020",
    number = "2020"
}

@article{Kierkla:2022odc,
    author = "Kierkla, Maciej and Karam, Alexandros and Swiezewska, Bogumila",
    title = "{Conformal model for gravitational waves and dark matter: a status update}",
    eprint = "2210.07075",
    archivePrefix = "arXiv",
    primaryClass = "astro-ph.CO",
    doi = "10.1007/JHEP03(2023)007",
    journal = "J. High Energy Phys.",
    volume = "03",
    pages = "007",
    year = "2023",
    number = "2023"
}

@article{Kamionkowski:1993fg,
    author = "Kamionkowski, Marc and Kosowsky, Arthur and Turner, Michael S.",
    title = "{Gravitational radiation from first order phase transitions}",
    eprint = "astro-ph/9310044",
    archivePrefix = "arXiv",
    reportNumber = "IASSNS-HEP-93-44, FERMILAB-PUB-93-235-A",
    doi = "10.1103/PhysRevD.49.2837",
    journal = "Phys. Rev. D",
    volume = "49",
    pages = "2837--2851",
    year = "1994"
}

@article{Jinno:2016knw,
    author = "Jinno, Ryusuke and Takimoto, Masahiro",
    title = "{Probing a classically conformal B-L model with gravitational waves}",
    eprint = "1604.05035",
    archivePrefix = "arXiv",
    primaryClass = "hep-ph",
    reportNumber = "KEK-TH-1896",
    doi = "10.1103/PhysRevD.95.015020",
    journal = "Phys. Rev. D",
    volume = "95",
    number = "1",
    pages = "015020",
    year = "2017"
}

@article{Iso:2017uuu,
    author = "Iso, Satoshi and Serpico, Pasquale D. and Shimada, Kengo",
    title = "{QCD-Electroweak First-Order Phase Transition in a Supercooled Universe}",
    eprint = "1704.04955",
    archivePrefix = "arXiv",
    primaryClass = "hep-ph",
    reportNumber = "KEK-TH-1969, LAPTH-008-17",
    doi = "10.1103/PhysRevLett.119.141301",
    journal = "Phys. Rev. Lett.",
    volume = "119",
    number = "14",
    pages = "141301",
    year = "2017"
}

@article{Marzola:2017jzl,
    author = "Marzola, Luca and Racioppi, Antonio and Vaskonen, Ville",
    title = "{Phase transition and gravitational wave phenomenology of scalar conformal extensions of the Standard Model}",
    eprint = "1704.01034",
    archivePrefix = "arXiv",
    primaryClass = "hep-ph",
    doi = "10.1140/epjc/s10052-017-4996-1",
    journal = "Eur. Phys. J. C",
    volume = "77",
    number = "7",
    pages = "484",
    year = "2017"
}

@article{Azatov:2019png,
    author = "Azatov, Aleksandr and Barducci, Daniele and Sgarlata, Francesco",
    title = "{Gravitational traces of broken gauge symmetries}",
    eprint = "1910.01124",
    archivePrefix = "arXiv",
    primaryClass = "hep-ph",
    reportNumber = "SISSA 28/2019/FISI",
    doi = "10.1088/1475-7516/2020/07/027",
    journal = "J. Cosmol. Astropart. Phys.",
    volume = "07",
    pages = "027",
    year = "2020",
    number = "2020"
}

@article{Borah:2021ftr,
    author = "Borah, Debasish and Dasgupta, Arnab and Kang, Sin Kyu",
    title = "{A first order dark SU(2)$_{D}$ phase transition with vector dark matter in the light of NANOGrav 12.5 yr data}",
    eprint = "2109.11558",
    archivePrefix = "arXiv",
    primaryClass = "hep-ph",
    doi = "10.1088/1475-7516/2021/12/039",
    journal = "JCAP",
    volume = "12",
    number = "12",
    pages = "039",
    year = "2021"
}

@article{Athron:2023xlk,
    author = "Athron, Peter and Bal{\'a}zs, Csaba and Fowlie, Andrew and Morris, Lachlan and Wu, Lei",
    title = "{Cosmological phase transitions: From perturbative particle physics to gravitational waves}",
    eprint = "2305.02357",
    archivePrefix = "arXiv",
    primaryClass = "hep-ph",
    doi = "10.1016/j.ppnp.2023.104094",
    journal = "Prog. Part. Nucl. Phys.",
    volume = "135",
    pages = "104094",
    year = "2024"
}

@article{Gould:2021ccf,
    author = "Gould, Oliver and Hirvonen, Joonas",
    title = "{Effective field theory approach to thermal bubble nucleation}",
    eprint = "2108.04377",
    archivePrefix = "arXiv",
    primaryClass = "hep-ph",
    reportNumber = "HIP-2020-19/TH",
    doi = "10.1103/PhysRevD.104.096015",
    journal = "Phys. Rev. D",
    volume = "104",
    number = "9",
    pages = "096015",
    year = "2021"
}

@article{Lofgren:2021ogg,
    author = {L{\"o}fgren, Johan and Ramsey-Musolf, Michael J. and Schicho, Philipp and Tenkanen, Tuomas V. I.},
    title = "{Nucleation at Finite Temperature: A Gauge-Invariant Perturbative Framework}",
    eprint = "2112.05472",
    archivePrefix = "arXiv",
    primaryClass = "hep-ph",
    reportNumber = "ACFI-T21-15, HIP-2021-44/TH, NORDITA 2021-110",
    doi = "10.1103/PhysRevLett.130.251801",
    journal = "Phys. Rev. Lett.",
    volume = "130",
    number = "25",
    pages = "251801",
    year = "2023"
}

@article{Hirvonen:2021zej,
    author = {Hirvonen, Joonas and L{\"o}fgren, Johan and Ramsey-Musolf, Michael J. and Schicho, Philipp and Tenkanen, Tuomas V. I.},
    title = "{Computing the gauge-invariant bubble nucleation rate in finite temperature effective field theory}",
    eprint = "2112.08912",
    archivePrefix = "arXiv",
    primaryClass = "hep-ph",
    reportNumber = "ACFI-T21-16, HIP-2021-45/TH, NORDITA 2021-111",
    doi = "10.1007/JHEP07(2022)135",
    journal = "J. High Energy Phys.",
    volume = "07",
    pages = "135",
    year = "2022",
    number = "2022"
}

@article{Schicho:2022wty,
    author = "Schicho, Philipp and Tenkanen, Tuomas V. I. and White, Graham",
    title = "{Combining thermal resummation and gauge invariance for electroweak phase transition}",
    eprint = "2203.04284",
    archivePrefix = "arXiv",
    primaryClass = "hep-ph",
    reportNumber = "HIP-2022-2/TH, NORDITA 2022-009",
    doi = "10.1007/JHEP11(2022)047",
    journal = "J. High Energy Phys.",
    volume = "11",
    pages = "047",
    year = "2022",
    number = "2022"
}

@article{Ekstedt:2022zro,
    author = {Ekstedt, Andreas and Gould, Oliver and L{\"o}fgren, Johan},
    title = "{Radiative first-order phase transitions to next-to-next-to-leading order}",
    eprint = "2205.07241",
    archivePrefix = "arXiv",
    primaryClass = "hep-ph",
    doi = "10.1103/PhysRevD.106.036012",
    journal = "Phys. Rev. D",
    volume = "106",
    number = "3",
    pages = "036012",
    year = "2022",
    note = "[Erratum: Phys.Rev.D 110, 019901 (2024)]"
}

@article{Lofgren:2023sep,
    author = {L{\"o}fgren, Johan},
    title = "{Stop comparing resummation methods}",
    eprint = "2301.05197",
    archivePrefix = "arXiv",
    primaryClass = "hep-ph",
    doi = "10.1088/1361-6471/ad074b",
    journal = "J. Phys. G",
    volume = "50",
    number = "12",
    pages = "125008",
    year = "2023"
}

@article{Gould:2023ovu,
    author = "Gould, Oliver and Tenkanen, Tuomas V. I.",
    title = "{Perturbative effective field theory expansions for cosmological phase transitions}",
    eprint = "2309.01672",
    archivePrefix = "arXiv",
    primaryClass = "hep-ph",
    reportNumber = "NORDITA 2023-037",
    doi = "10.1007/JHEP01(2024)048",
    journal = "J. High Energy Phys.",
    volume = "01",
    pages = "048",
    year = "2024",
    number = "2024"
}

@article{Parwani:1991gq,
    author = "Parwani, Rajesh R.",
    title = "{Resummation in a hot scalar field theory}",
    eprint = "hep-ph/9204216",
    archivePrefix = "arXiv",
    reportNumber = "ITP-SB-91-64",
    doi = "10.1103/PhysRevD.45.4695",
    journal = "Phys. Rev. D",
    volume = "45",
    pages = "4695",
    year = "1992",
    note = "[Erratum: Phys.Rev.D 48, 5965 (1993)]"
}

@article{Brazier:2019mmu,
    author = "Ransom, Scott and others",
    collaboration = "NANOGrav",
    title = "{The NANOGrav Program for Gravitational Waves and Fundamental Physics}",
    eprint = "1908.05356",
    archivePrefix = "arXiv",
    primaryClass = "astro-ph.IM",
    month = "8",
    year = "2019",
    journal = "Bull. Am. Astron. Soc.",
    number = "7",
    volume = "51"
}

@article{Manchester:2012za,
    author = "Manchester, R. N. and others",
    title = "{The Parkes Pulsar Timing Array Project}",
    eprint = "1210.6130",
    archivePrefix = "arXiv",
    primaryClass = "astro-ph.IM",
    doi = "10.1017/pasa.2012.017",
    journal = "Publ. Astron. Soc. Austral.",
    volume = "30",
    pages = "17",
    year = "2013"
}

@article{ChandraJoshi:2022etw,
    author = "Chandra Joshi, Bhal and others",
    title = "{Nanohertz gravitational wave astronomy during SKA era: An InPTA perspective}",
    eprint = "2207.06461",
    archivePrefix = "arXiv",
    primaryClass = "astro-ph.HE",
    doi = "10.1007/s12036-022-09869-w",
    journal = "J. Astrophys. Astron.",
    volume = "43",
    number = "2",
    pages = "98",
    year = "2022"
}

@INPROCEEDINGS{Lee:2016xxx,
       author = {{Lee}, K.~J.},
        title = "{Prospects of Gravitational Wave Detection Using Pulsar Timing Array for Chinese Future Telescopes}",
    booktitle = {Frontiers in Radio Astronomy and FAST Early Sciences Symposium 2015},
         year = 2016,
       editor = {{Qain}, L. and {Li}, D.},
       series = {Astronomical Society of the Pacific Conference Series},
       volume = {502},
        month = feb,
        pages = {19},
       adsurl = {https://ui.adsabs.harvard.edu/abs/2016ASPC..502...19L}
}

@article{Miles:2022lkg,
    author = "Miles, Matthew T. and others",
    title = "{The MeerKAT Pulsar Timing Array: first data release}",
    eprint = "2212.04648",
    archivePrefix = "arXiv",
    primaryClass = "astro-ph.HE",
    doi = "10.1093/mnras/stac3644",
    journal = "Mon. Not. Roy. Astron. Soc.",
    volume = "519",
    number = "3",
    pages = "3976--3991",
    year = "2023"
}

@article{Manchester:2013ndt,
    author = "Manchester, R. N.",
    title = "{The International Pulsar Timing Array}",
    eprint = "1309.7392",
    archivePrefix = "arXiv",
    primaryClass = "astro-ph.IM",
    doi = "10.1088/0264-9381/30/22/224010",
    journal = "Class. Quant. Grav.",
    volume = "30",
    pages = "224010",
    year = "2013"
}

@article{Verbiest:2009kb,
    author = "Verbiest, J. P. W. and others",
    title = "{Timing stability of millisecond pulsars and prospects for gravitational-wave detection}",
    eprint = "0908.0244",
    archivePrefix = "arXiv",
    primaryClass = "astro-ph.GA",
    doi = "10.1111/j.1365-2966.2009.15508.x",
    journal = "Mon. Not. Roy. Astron. Soc.",
    volume = "400",
    pages = "951--968",
    year = "2009"
}

@article{EPTA:2023fyk,
    author = "Antoniadis, J. and others",
    collaboration = "EPTA, InPTA:",
    title = "{The second data release from the European Pulsar Timing Array - III. Search for gravitational wave signals}",
    eprint = "2306.16214",
    archivePrefix = "arXiv",
    primaryClass = "astro-ph.HE",
    doi = "10.1051/0004-6361/202346844",
    journal = "Astron. Astrophys.",
    volume = "678",
    pages = "A50",
    year = "2023"
}

@article{Reardon:2023gzh,
    author = "Reardon, Daniel J. and others",
    title = "{Search for an Isotropic Gravitational-wave Background with the Parkes Pulsar Timing Array}",
    eprint = "2306.16215",
    archivePrefix = "arXiv",
    primaryClass = "astro-ph.HE",
    doi = "10.3847/2041-8213/acdd02",
    journal = "Astrophys. J. Lett.",
    volume = "951",
    number = "1",
    pages = "L6",
    year = "2023"
}

@article{Xu:2023wog,
    author = "Xu, Heng and others",
    title = "{Searching for the Nano-Hertz Stochastic Gravitational Wave Background with the Chinese Pulsar Timing Array Data Release I}",
    eprint = "2306.16216",
    archivePrefix = "arXiv",
    primaryClass = "astro-ph.HE",
    doi = "10.1088/1674-4527/acdfa5",
    journal = "Res. Astron. Astrophys.",
    volume = "23",
    number = "7",
    pages = "075024",
    year = "2023"
}

@article{Miles:2024seg,
    author = "Miles, Matthew T. and others",
    title = "{The MeerKAT Pulsar Timing Array: the first search for gravitational waves with the MeerKAT radio telescope}",
    eprint = "2412.01153",
    archivePrefix = "arXiv",
    primaryClass = "astro-ph.HE",
    doi = "10.1093/mnras/stae2571",
    journal = "Mon. Not. Roy. Astron. Soc.",
    volume = "536",
    number = "2",
    pages = "1489--1500",
    year = "2024"
}

@article{NANOGrav:2023gor,
    author = "Agazie, Gabriella and others",
    collaboration = "NANOGrav",
    title = "{The NANOGrav 15 yr Data Set: Evidence for a Gravitational-wave Background}",
    eprint = "2306.16213",
    archivePrefix = "arXiv",
    primaryClass = "astro-ph.HE",
    doi = "10.3847/2041-8213/acdac6",
    journal = "Astrophys. J. Lett.",
    volume = "951",
    number = "1",
    pages = "L8",
    year = "2023"
}

@article{Rajagopal:1994zj,
    author = "Rajagopal, Mohan and Romani, Roger W.",
    title = "{Ultralow frequency gravitational radiation from massive black hole binaries}",
    eprint = "astro-ph/9412038",
    archivePrefix = "arXiv",
    doi = "10.1086/175813",
    journal = "Astrophys. J.",
    volume = "446",
    pages = "543--549",
    year = "1995"
}

@article{Jaffe:2002rt,
    author = "Jaffe, Andrew H. and Backer, Donald C.",
    title = "{Gravitational waves probe the coalescence rate of massive black hole binaries}",
    eprint = "astro-ph/0210148",
    archivePrefix = "arXiv",
    doi = "10.1086/345443",
    journal = "Astrophys. J.",
    volume = "583",
    pages = "616--631",
    year = "2003"
}

@article{Wyithe:2002ep,
    author = "Wyithe, J. Stuart B. and Loeb, Abraham",
    title = "{Low - frequency gravitational waves from massive black hole binaries: Predictions for LISA and pulsar timing arrays}",
    eprint = "astro-ph/0211556",
    archivePrefix = "arXiv",
    doi = "10.1086/375187",
    journal = "Astrophys. J.",
    volume = "590",
    pages = "691--706",
    year = "2003"
}

@article{Sesana:2004sp,
    author = "Sesana, Alberto and Haardt, Francesco and Madau, Piero and Volonteri, Marta",
    title = "{Low - frequency gravitational radiation from coalescing massive black hole binaries in hierarchical cosmologies}",
    eprint = "astro-ph/0401543",
    archivePrefix = "arXiv",
    doi = "10.1086/422185",
    journal = "Astrophys. J.",
    volume = "611",
    pages = "623--632",
    year = "2004"
}

@article{McWilliams:2012an,
    author = "McWilliams, Sean T. and Ostriker, Jeremiah P. and Pretorius, Frans",
    title = "{Gravitational waves and stalled satellites from massive galaxy mergers at $z \leq 1$}",
    eprint = "1211.5377",
    archivePrefix = "arXiv",
    primaryClass = "astro-ph.CO",
    doi = "10.1088/0004-637X/789/2/156",
    journal = "Astrophys. J.",
    volume = "789",
    pages = "156",
    year = "2014"
}

@article{Burke-Spolaor:2018bvk,
    author = "Burke-Spolaor, Sarah and others",
    title = "{The Astrophysics of Nanohertz Gravitational Waves}",
    eprint = "1811.08826",
    archivePrefix = "arXiv",
    primaryClass = "astro-ph.HE",
    doi = "10.1007/s00159-019-0115-7",
    journal = "Astron. Astrophys. Rev.",
    volume = "27",
    number = "1",
    pages = "5",
    year = "2019"
}

@article{Witten:1984rs,
    author = "Witten, Edward",
    title = "{Cosmic Separation of Phases}",
    reportNumber = "PRINT-84-0400 (IAS,PRINCETON)",
    doi = "10.1103/PhysRevD.30.272",
    journal = "Phys. Rev. D",
    volume = "30",
    pages = "272--285",
    year = "1984"
}

@article{Hogan:1986dsh,
    author = "Hogan, C. J.",
    title = "{Gravitational radiation from cosmological phase transitions}",
    doi = "10.1093/mnras/218.4.629",
    journal = "Mon. Not. Roy. Astron. Soc.",
    volume = "218",
    number = "4",
    pages = "629--636",
    year = "1986"
}

@article{Athron:2023mer,
    author = "Athron, Peter and Fowlie, Andrew and Lu, Chih-Ting and Morris, Lachlan and Wu, Lei and Wu, Yongcheng and Xu, Zhongxiu",
    title = "{Can Supercooled Phase Transitions Explain the Gravitational Wave Background Observed by Pulsar Timing Arrays?}",
    eprint = "2306.17239",
    archivePrefix = "arXiv",
    primaryClass = "hep-ph",
    doi = "10.1103/PhysRevLett.132.221001",
    journal = "Phys. Rev. Lett.",
    volume = "132",
    number = "22",
    pages = "221001",
    year = "2024"
}

@article{Goncalves:2025uwh,
    author = "Gon{\c{c}}alves, Jo{\~a}o and Marfatia, Danny and Morais, Ant{\'o}nio P. and Pasechnik, Roman",
    title = "{Supercooled phase transitions in conformal dark sectors explain NANOGrav data}",
    eprint = "2501.11619",
    archivePrefix = "arXiv",
    primaryClass = "hep-ph",
    doi = "10.1016/j.physletb.2025.139829",
    journal = "Phys. Lett. B",
    volume = "869",
    pages = "139829",
    year = "2025"
}

@article{Costa:2025csj,
    author = "Costa, Francesco and Hoefken Zink, Jaime and Lucente, Michele and Pascoli, Silvia and Rosauro-Alcaraz, Salvador",
    title = "{Supercooled dark scalar phase transitions explanation of NANOGrav data}",
    eprint = "2501.15649",
    archivePrefix = "arXiv",
    primaryClass = "hep-ph",
    doi = "10.1016/j.physletb.2025.139634",
    journal = "Phys. Lett. B",
    volume = "868",
    pages = "139634",
    year = "2025"
}

@article{Weinberg:1974hy,
    author = "Weinberg, Steven",
    title = "{Gauge and Global Symmetries at High Temperature}",
    reportNumber = "PRINT-74-0689 (HARVARD)",
    doi = "10.1103/PhysRevD.9.3357",
    journal = "Phys. Rev. D",
    volume = "9",
    pages = "3357--3378",
    year = "1974"
}

@article{Dolan:1973qd,
    author = "Dolan, L. and Jackiw, R.",
    title = "{Symmetry Behavior at Finite Temperature}",
    reportNumber = "MIT-CTP-406",
    doi = "10.1103/PhysRevD.9.3320",
    journal = "Phys. Rev. D",
    volume = "9",
    pages = "3320--3341",
    year = "1974"
}

@article{Linde:1978px,
    author = "Linde, Andrei D.",
    title = "{Phase Transitions in Gauge Theories and Cosmology}",
    reportNumber = "LEBEDEV-78-166",
    doi = "10.1088/0034-4885/42/3/001",
    journal = "Rept. Prog. Phys.",
    volume = "42",
    pages = "389",
    year = "1979"
}

@article{Linde:1980ts,
    author = "Linde, Andrei D.",
    title = "{Infrared Problem in Thermodynamics of the Yang-Mills Gas}",
    reportNumber = "LEBEDEV-80-106",
    doi = "10.1016/0370-2693(80)90769-8",
    journal = "Phys. Lett. B",
    volume = "96",
    pages = "289--292",
    year = "1980"
}

@article{Farakos:1994kx,
    author = "Farakos, K. and Kajantie, K. and Rummukainen, K. and Shaposhnikov, Mikhail E.",
    title = "{3-D physics and the electroweak phase transition: Perturbation theory}",
    eprint = "hep-ph/9404201",
    archivePrefix = "arXiv",
    reportNumber = "CERN-TH-6973-94, IUHET-273",
    doi = "10.1016/0550-3213(94)90173-2",
    journal = "Nucl. Phys. B",
    volume = "425",
    pages = "67--109",
    year = "1994"
}

@article{Braaten:1995cm,
    author = "Braaten, Eric and Nieto, Agustin",
    title = "{Effective field theory approach to high temperature thermodynamics}",
    eprint = "hep-ph/9501375",
    archivePrefix = "arXiv",
    reportNumber = "NUHEP-TH-95-2",
    doi = "10.1103/PhysRevD.51.6990",
    journal = "Phys. Rev. D",
    volume = "51",
    pages = "6990--7006",
    year = "1995"
}

@article{Kajantie:1995dw,
    author = "Kajantie, K. and Laine, M. and Rummukainen, K. and Shaposhnikov, Mikhail E.",
    title = "{Generic rules for high temperature dimensional reduction and their application to the standard model}",
    eprint = "hep-ph/9508379",
    archivePrefix = "arXiv",
    reportNumber = "CERN-TH-95-226, HU-TFT-95-50, IUHET-312",
    doi = "10.1016/0550-3213(95)00549-8",
    journal = "Nucl. Phys. B",
    volume = "458",
    pages = "90--136",
    year = "1996"
}

@article{Boyd:1993tz,
    author = "Boyd, C. Glenn and Brahm, David E. and Hsu, Stephen D. H.",
    title = "{Resummation methods at finite temperature: The Tadpole way}",
    eprint = "hep-ph/9304254",
    archivePrefix = "arXiv",
    reportNumber = "CALT-68-1858, HUTP-93-A011, EFI-93-22",
    doi = "10.1103/PhysRevD.48.4963",
    journal = "Phys. Rev. D",
    volume = "48",
    pages = "4963--4973",
    year = "1993"
}

@article{Curtin:2016urg,
    author = "Curtin, David and Meade, Patrick and Ramani, Harikrishnan",
    title = "{Thermal Resummation and Phase Transitions}",
    eprint = "1612.00466",
    archivePrefix = "arXiv",
    primaryClass = "hep-ph",
    reportNumber = "YITP-2016-48",
    doi = "10.1140/epjc/s10052-018-6268-0",
    journal = "Eur. Phys. J. C",
    volume = "78",
    number = "9",
    pages = "787",
    year = "2018"
}

@article{Curtin:2022ovx,
    author = "Curtin, David and Roy, Jyotirmoy and White, Graham",
    title = "{Gravitational waves and tadpole resummation: Efficient and easy convergence of finite temperature QFT}",
    eprint = "2211.08218",
    archivePrefix = "arXiv",
    primaryClass = "hep-ph",
    doi = "10.1103/PhysRevD.109.116001",
    journal = "Phys. Rev. D",
    volume = "109",
    number = "11",
    pages = "116001",
    year = "2024"
}

@article{Bahl:2024ykv,
    author = "Bahl, Henning and Carena, Marcela and Ireland, Aurora and Wagner, Carlos E. M.",
    title = "{Improved thermal resummation for multi-field potentials}",
    eprint = "2404.12439",
    archivePrefix = "arXiv",
    primaryClass = "hep-ph",
    reportNumber = "EFI 24-2, FERMILAB-PUB-24-0040-T",
    doi = "10.1007/JHEP09(2024)153",
    journal = "J. High Energy Phys.",
    volume = "09",
    pages = "153",
    year = "2024",
    number = "2024"
}

@article{Bittar:2025lcr,
    author = "Bittar, Pedro and Roy, Subhojit and Wagner, Carlos E. M.",
    title = "{Self consistent thermal resummation: a case study of the phase transition in 2HDM}",
    eprint = "2504.02024",
    archivePrefix = "arXiv",
    primaryClass = "hep-ph",
    reportNumber = "EFI-25-3",
    doi = "10.1007/JHEP12(2025)021",
    journal = "J. High Energy Phys.",
    volume = "12",
    pages = "021",
    year = "2025",
    number = "2025"
}

@article{Kierkla:2025qyz,
    author = "Kierkla, Maciej and Schicho, Philipp and Swiezewska, Bogumila and Tenkanen, Tuomas V. I. and van de Vis, Jorinde",
    title = "{Finite-temperature bubble nucleation with shifting scale hierarchies}",
    eprint = "2503.13597",
    archivePrefix = "arXiv",
    primaryClass = "hep-ph",
    reportNumber = "CERN-TH-2025-046, HIP-2024-27/TH",
    doi = "10.1007/JHEP07(2025)153",
    journal = "J. High Energy Phys.",
    volume = "07",
    pages = "153",
    year = "2025",
    number = "2025"
}

@article{Kosowsky:1991ua,
    author = "Kosowsky, Arthur and Turner, Michael S. and Watkins, Richard",
    title = "{Gravitational radiation from colliding vacuum bubbles}",
    reportNumber = "FERMILAB-PUB-91-323-A",
    doi = "10.1103/PhysRevD.45.4514",
    journal = "Phys. Rev. D",
    volume = "45",
    pages = "4514--4535",
    year = "1992"
}

@article{Vachaspati:1984gt,
    author = "Vachaspati, Tanmay and Vilenkin, Alexander",
    title = "{Gravitational Radiation from Cosmic Strings}",
    reportNumber = "HUTP-84/A065",
    doi = "10.1103/PhysRevD.31.3052",
    journal = "Phys. Rev. D",
    volume = "31",
    pages = "3052",
    year = "1985"
}

@article{Sakellariadou:1990ne,
    author = "Sakellariadou, M.",
    title = "{Gravitational waves emitted from infinite strings}",
    doi = "10.1103/PhysRevD.42.354",
    journal = "Phys. Rev. D",
    volume = "42",
    pages = "354--360",
    year = "1990",
    note = "[Erratum: Phys.Rev.D 43, 4150 (1991)]"
}

@article{Hiramatsu:2013qaa,
    author = "Hiramatsu, Takashi and Kawasaki, Masahiro and Saikawa, Ken'ichi",
    title = "{On the estimation of gravitational wave spectrum from cosmic domain walls}",
    eprint = "1309.5001",
    archivePrefix = "arXiv",
    primaryClass = "astro-ph.CO",
    reportNumber = "ICRR-REPORT-659-2013-8, IPMU13-0182, YITP-13-87",
    doi = "10.1088/1475-7516/2014/02/031",
    journal = "J. Cosmol. Astropart. Phys.",
    volume = "02",
    pages = "031",
    year = "2014",
    number = "2014"
}

@article{Turner:1996ck,
    author = "Turner, Michael S.",
    title = "{Detectability of inflation produced gravitational waves}",
    eprint = "astro-ph/9607066",
    archivePrefix = "arXiv",
    reportNumber = "FERMILAB-PUB-96-169-A, FERMILAB-PUB-96-167-A",
    doi = "10.1103/PhysRevD.55.R435",
    journal = "Phys. Rev. D",
    volume = "55",
    pages = "R435--R439",
    year = "1997"
}

@article{Bodeker:2009qy,
    author = "Bodeker, Dietrich and Moore, Guy D.",
    title = "{Can electroweak bubble walls run away?}",
    eprint = "0903.4099",
    archivePrefix = "arXiv",
    primaryClass = "hep-ph",
    doi = "10.1088/1475-7516/2009/05/009",
    journal = "J. Cosmol. Astropart. Phys.",
    volume = "05",
    pages = "009",
    year = "2009",
    number = "2009"
}

@article{Bodeker:2017cim,
    author = "Bodeker, Dietrich and Moore, Guy D.",
    title = "{Electroweak Bubble Wall Speed Limit}",
    eprint = "1703.08215",
    archivePrefix = "arXiv",
    primaryClass = "hep-ph",
    doi = "10.1088/1475-7516/2017/05/025",
    journal = "J. Cosmol. Astropart. Phys.",
    volume = "05",
    pages = "025",
    year = "2017",
    number = "2017"
}

@article{Konstandin:2011dr,
    author = "Konstandin, Thomas and Servant, Geraldine",
    title = "{Cosmological Consequences of Nearly Conformal Dynamics at the TeV scale}",
    eprint = "1104.4791",
    archivePrefix = "arXiv",
    primaryClass = "hep-ph",
    doi = "10.1088/1475-7516/2011/12/009",
    journal = "J. Cosmol. Astropart. Phys.",
    volume = "12",
    pages = "009",
    year = "2011",
    number = "2011"
}

@article{Konstandin:2011ds,
    author = "Konstandin, Thomas and Servant, Geraldine",
    title = "{Natural Cold Baryogenesis from Strongly Interacting Electroweak Symmetry Breaking}",
    eprint = "1104.4793",
    archivePrefix = "arXiv",
    primaryClass = "hep-ph",
    doi = "10.1088/1475-7516/2011/07/024",
    journal = "J. Cosmol. Astropart. Phys.",
    volume = "07",
    pages = "024",
    year = "2011",
    number = "2011"
}

@article{vonHarling:2017yew,
    author = "von Harling, Benedict and Servant, Geraldine",
    title = "{QCD-induced Electroweak Phase Transition}",
    eprint = "1711.11554",
    archivePrefix = "arXiv",
    primaryClass = "hep-ph",
    reportNumber = "DESY-17-056",
    doi = "10.1007/JHEP01(2018)159",
    journal = "J. High Energy Phys.",
    volume = "01",
    pages = "159",
    year = "2018",
    number = "2018"
}

@article{Prokopec:2018tnq,
    author = "Prokopec, Tomislav and Rezacek, Jonas and {\'S}wie{\.z}ewska, Bogumi{\l}a",
    title = "{Gravitational waves from conformal symmetry breaking}",
    eprint = "1809.11129",
    archivePrefix = "arXiv",
    primaryClass = "hep-ph",
    doi = "10.1088/1475-7516/2019/02/009",
    journal = "J. Cosmol. Astropart. Phys.",
    volume = "02",
    pages = "009",
    year = "2019",
    number = "2019"
}

@article{Baratella:2018pxi,
    author = "Baratella, Pietro and Pomarol, Alex and Rompineve, Fabrizio",
    title = "{The Supercooled Universe}",
    eprint = "1812.06996",
    archivePrefix = "arXiv",
    primaryClass = "hep-ph",
    doi = "10.1007/JHEP03(2019)100",
    journal = "J. High Energy Phys.",
    volume = "03",
    pages = "100",
    year = "2019",
    number = "2019"
}

@article{Athron:2022mmm,
    author = "Athron, Peter and Bal{\'a}zs, Csaba and Morris, Lachlan",
    title = "{Supercool subtleties of cosmological phase transitions}",
    eprint = "2212.07559",
    archivePrefix = "arXiv",
    primaryClass = "hep-ph",
    doi = "10.1088/1475-7516/2023/03/006",
    journal = "J. Cosmol. Astropart. Phys.",
    volume = "03",
    pages = "006",
    year = "2023",
    number = "2023"
}

@article{Ramberg:2022irf,
    author = "Ramberg, Nicklas and Ratzinger, Wolfram and Schwaller, Pedro",
    title = "{One {\ensuremath{\mu}} to rule them all: CMB spectral distortions can probe domain walls, cosmic strings and low scale phase transitions}",
    eprint = "2209.14313",
    archivePrefix = "arXiv",
    primaryClass = "hep-ph",
    reportNumber = "MITP-22-077",
    doi = "10.1088/1475-7516/2023/02/039",
    journal = "J. Cosmol. Astropart. Phys.",
    volume = "02",
    pages = "039",
    year = "2023",
    number = "2023"
}

@article{Bringmann:2025cht,
    author = "Bringmann, Torsten and Croon, Djuna and Sevillano Mu{\~n}oz, Sergio",
    title = "{Updated constraints on the primordial power spectrum at sub-Mpc scales}",
    eprint = "2506.20704",
    archivePrefix = "arXiv",
    primaryClass = "astro-ph.CO",
    reportNumber = "IPPP/25/40",
    month = "6",
    year = "2025",
}

@book{Laine:2016hma,
    author = "Laine, Mikko and Vuorinen, Aleksi",
    title = "{Basics of Thermal Field Theory}",
    eprint = "1701.01554",
    archivePrefix = "arXiv",
    primaryClass = "hep-ph",
    doi = "10.1007/978-3-319-31933-9",
    publisher = "Springer",
    volume = "925",
    year = "2016"
}

@article{Jinno:2017fby,
    author = "Jinno, Ryusuke and Takimoto, Masahiro",
    title = "{Gravitational waves from bubble dynamics: Beyond the Envelope}",
    eprint = "1707.03111",
    archivePrefix = "arXiv",
    primaryClass = "hep-ph",
    reportNumber = "CTPU-17-26, KEK-TH-1986",
    doi = "10.1088/1475-7516/2019/01/060",
    journal = "J. Cosmol. Astropart. Phys.",
    volume = "01",
    pages = "060",
    year = "2019",
    number = "2019"
}

@article{Konstandin:2017sat,
    author = "Konstandin, Thomas",
    title = "{Gravitational radiation from a bulk flow model}",
    eprint = "1712.06869",
    archivePrefix = "arXiv",
    primaryClass = "astro-ph.CO",
    reportNumber = "DESY-17-227",
    doi = "10.1088/1475-7516/2018/03/047",
    journal = "J. Cosmol. Astropart. Phys.",
    volume = "03",
    pages = "047",
    year = "2018",
    number = "2018"
}

@article{Ellis:2020ena,
    author = "Ellis, John and Lewicki, Marek",
    title = "{Cosmic String Interpretation of NANOGrav Pulsar Timing Data}",
    eprint = "2009.06555",
    archivePrefix = "arXiv",
    primaryClass = "astro-ph.CO",
    reportNumber = "KCL-PH-TH/2020-53, CERN-TH-2020-150",
    doi = "10.1103/PhysRevLett.126.041304",
    journal = "Phys. Rev. Lett.",
    volume = "126",
    number = "4",
    pages = "041304",
    year = "2021"
}

@article{Blasi:2020mfx,
    author = "Blasi, Simone and Brdar, Vedran and Schmitz, Kai",
    title = "{Has NANOGrav found first evidence for cosmic strings?}",
    eprint = "2009.06607",
    archivePrefix = "arXiv",
    primaryClass = "astro-ph.CO",
    reportNumber = "CERN-TH-2020-151",
    doi = "10.1103/PhysRevLett.126.041305",
    journal = "Phys. Rev. Lett.",
    volume = "126",
    number = "4",
    pages = "041305",
    year = "2021"
}

@article{DeLuca:2020agl,
    author = "De Luca, V. and Franciolini, G. and Riotto, A.",
    title = "{NANOGrav Data Hints at Primordial Black Holes as Dark Matter}",
    eprint = "2009.08268",
    archivePrefix = "arXiv",
    primaryClass = "astro-ph.CO",
    doi = "10.1103/PhysRevLett.126.041303",
    journal = "Phys. Rev. Lett.",
    volume = "126",
    number = "4",
    pages = "041303",
    year = "2021"
}

@article{Buchmuller:2020lbh,
    author = "Buchmuller, Wilfried and Domcke, Valerie and Schmitz, Kai",
    title = "{From NANOGrav to LIGO with metastable cosmic strings}",
    eprint = "2009.10649",
    archivePrefix = "arXiv",
    primaryClass = "astro-ph.CO",
    reportNumber = "CERN-TH-2020-157, DESY 20-154, DESY-20-154",
    doi = "10.1016/j.physletb.2020.135914",
    journal = "Phys. Lett. B",
    volume = "811",
    pages = "135914",
    year = "2020"
}

@article{Addazi:2020zcj,
    author = "Addazi, Andrea and Cai, Yi-Fu and Gan, Qingyu and Marciano, Antonino and Zeng, Kaiqiang",
    title = "{NANOGrav results and dark first order phase transitions}",
    eprint = "2009.10327",
    archivePrefix = "arXiv",
    primaryClass = "hep-ph",
    doi = "10.1007/s11433-021-1724-6",
    journal = "Sci. China Phys. Mech. Astron.",
    volume = "64",
    number = "9",
    pages = "290411",
    year = "2021"
}

@article{Vagnozzi:2020gtf,
    author = "Vagnozzi, Sunny",
    title = "{Implications of the NANOGrav results for inflation}",
    eprint = "2009.13432",
    archivePrefix = "arXiv",
    primaryClass = "astro-ph.CO",
    doi = "10.1093/mnrasl/slaa203",
    journal = "Mon. Not. Roy. Astron. Soc.",
    volume = "502",
    number = "1",
    pages = "L11--L15",
    year = "2021"
}

@article{Bian:2020urb,
    author = "Bian, Ligong and Cai, Rong-Gen and Liu, Jing and Yang, Xing-Yu and Zhou, Ruiyu",
    title = "{Evidence for different gravitational-wave sources in the NANOGrav dataset}",
    eprint = "2009.13893",
    archivePrefix = "arXiv",
    primaryClass = "astro-ph.CO",
    doi = "10.1103/PhysRevD.103.L081301",
    journal = "Phys. Rev. D",
    volume = "103",
    number = "8",
    pages = "L081301",
    year = "2021"
}

@article{Neronov:2020qrl,
    author = "Neronov, Andrii and Roper Pol, Alberto and Caprini, Chiara and Semikoz, Dmitri",
    title = "{NANOGrav signal from magnetohydrodynamic turbulence at the QCD phase transition in the early Universe}",
    eprint = "2009.14174",
    archivePrefix = "arXiv",
    primaryClass = "astro-ph.CO",
    doi = "10.1103/PhysRevD.103.L041302",
    journal = "Phys. Rev. D",
    volume = "103",
    number = "4",
    pages = "041302",
    year = "2021"
}

@article{Li:2020cjj,
    author = "Li, Hao-Hao and Ye, Gen and Piao, Yun-Song",
    title = "{Is the NANOGrav signal a hint of dS decay during inflation?}",
    eprint = "2009.14663",
    archivePrefix = "arXiv",
    primaryClass = "astro-ph.CO",
    doi = "10.1016/j.physletb.2021.136211",
    journal = "Phys. Lett. B",
    volume = "816",
    pages = "136211",
    year = "2021"
}

@article{EuropeanPulsarTimingArray:2023lqe,
    author = "Quelquejay Leclere, Hippolyte and others",
    collaboration = "European Pulsar Timing Array, EPTA",
    title = "{Practical approaches to analyzing PTA data: Cosmic strings with six pulsars}",
    eprint = "2306.12234",
    archivePrefix = "arXiv",
    primaryClass = "gr-qc",
    doi = "10.1103/PhysRevD.108.123527",
    journal = "Phys. Rev. D",
    volume = "108",
    number = "12",
    pages = "123527",
    year = "2023"
}

@article{Franciolini:2023wjm,
    author = "Franciolini, Gabriele and Racco, Davide and Rompineve, Fabrizio",
    title = "{Footprints of the QCD Crossover on Cosmological Gravitational Waves at Pulsar Timing Arrays}",
    eprint = "2306.17136",
    archivePrefix = "arXiv",
    primaryClass = "astro-ph.CO",
    reportNumber = "CERN-TH-2023-080",
    doi = "10.1103/PhysRevLett.132.081001",
    journal = "Phys. Rev. Lett.",
    volume = "132",
    number = "8",
    pages = "081001",
    year = "2024",
    note = "[Erratum: Phys.Rev.Lett. 133, 189901 (2024)]"
}

@article{Guo:2023hyp,
    author = "Guo, Shu-Yuan and Khlopov, Maxim and Liu, Xuewen and Wu, Lei and Wu, Yongcheng and Zhu, Bin",
    title = "{Footprints of axion-like particle in pulsar timing array data and James Webb Space Telescope observations}",
    eprint = "2306.17022",
    archivePrefix = "arXiv",
    primaryClass = "hep-ph",
    doi = "10.1007/s11433-024-2445-1",
    journal = "Sci. China Phys. Mech. Astron.",
    volume = "67",
    number = "11",
    pages = "111011",
    year = "2024"
}

@article{Wang:2023len,
    author = "Wang, Ziwei and Lei, Lei and Jiao, Hao and Feng, Lei and Fan, Yi-Zhong",
    title = "{The nanohertz stochastic gravitational wave background from cosmic string loops and the abundant high redshift massive galaxies}",
    eprint = "2306.17150",
    archivePrefix = "arXiv",
    primaryClass = "astro-ph.HE",
    doi = "10.1007/s11433-023-2262-0",
    journal = "Sci. China Phys. Mech. Astron.",
    volume = "66",
    number = "12",
    pages = "120403",
    year = "2023"
}

@article{Ellis:2023tsl,
    author = "Ellis, John and Lewicki, Marek and Lin, Chunshan and Vaskonen, Ville",
    title = "{Cosmic superstrings revisited in light of NANOGrav 15-year data}",
    eprint = "2306.17147",
    archivePrefix = "arXiv",
    primaryClass = "astro-ph.CO",
    reportNumber = "KCL-PH-TH/2023-38, CERN-TH-2023-126, AION-REPORT/2023-07",
    doi = "10.1103/PhysRevD.108.103511",
    journal = "Phys. Rev. D",
    volume = "108",
    number = "10",
    pages = "103511",
    year = "2023"
}

@article{Vagnozzi:2023lwo,
    author = "Vagnozzi, Sunny",
    title = "{Inflationary interpretation of the stochastic gravitational wave background signal detected by pulsar timing array experiments}",
    eprint = "2306.16912",
    archivePrefix = "arXiv",
    primaryClass = "astro-ph.CO",
    doi = "10.1016/j.jheap.2023.07.001",
    journal = "JHEAp",
    volume = "39",
    pages = "81--98",
    year = "2023"
}

@article{Fujikura:2023lkn,
    author = "Fujikura, Kohei and Girmohanta, Sudhakantha and Nakai, Yuichiro and Suzuki, Motoo",
    title = "{NANOGrav signal from a dark conformal phase transition}",
    eprint = "2306.17086",
    archivePrefix = "arXiv",
    primaryClass = "hep-ph",
    reportNumber = "UT-Komaba/23-6",
    doi = "10.1016/j.physletb.2023.138203",
    journal = "Phys. Lett. B",
    volume = "846",
    pages = "138203",
    year = "2023"
}

@article{Kitajima:2023cek,
    author = "Kitajima, Naoya and Lee, Junseok and Murai, Kai and Takahashi, Fuminobu and Yin, Wen",
    title = "{Gravitational waves from domain wall collapse, and application to nanohertz signals with QCD-coupled axions}",
    eprint = "2306.17146",
    archivePrefix = "arXiv",
    primaryClass = "hep-ph",
    reportNumber = "TU-1198",
    doi = "10.1016/j.physletb.2024.138586",
    journal = "Phys. Lett. B",
    volume = "851",
    pages = "138586",
    year = "2024"
}

@article{Megias:2023kiy,
    author = "Megias, Eugenio and Nardini, Germano and Quiros, Mariano",
    title = "{Pulsar timing array stochastic background from light Kaluza-Klein resonances}",
    eprint = "2306.17071",
    archivePrefix = "arXiv",
    primaryClass = "hep-ph",
    doi = "10.1103/PhysRevD.108.095017",
    journal = "Phys. Rev. D",
    volume = "108",
    number = "9",
    pages = "095017",
    year = "2023"
}

@article{Bai:2023cqj,
    author = "Bai, Yang and Chen, Ting-Kuo and Korwar, Mrunal",
    title = "{QCD-collapsed domain walls: QCD phase transition and gravitational wave spectroscopy}",
    eprint = "2306.17160",
    archivePrefix = "arXiv",
    primaryClass = "hep-ph",
    doi = "10.1007/JHEP12(2023)194",
    journal = "J. High Energy Phys.",
    volume = "12",
    pages = "194",
    year = "2023",
    number = "2023"
}

@article{Addazi:2023jvg,
    author = "Addazi, Andrea and Cai, Yi-Fu and Marciano, Antonino and Visinelli, Luca",
    title = "{Have pulsar timing array methods detected a cosmological phase transition?}",
    eprint = "2306.17205",
    archivePrefix = "arXiv",
    primaryClass = "astro-ph.CO",
    reportNumber = "CA21106; CA21136",
    doi = "10.1103/PhysRevD.109.015028",
    journal = "Phys. Rev. D",
    volume = "109",
    number = "1",
    pages = "015028",
    year = "2024"
}

@article{Kitajima:2023vre,
    author = "Kitajima, Naoya and Nakayama, Kazunori",
    title = "{Nanohertz gravitational waves from cosmic strings and dark photon dark matter}",
    eprint = "2306.17390",
    archivePrefix = "arXiv",
    primaryClass = "hep-ph",
    reportNumber = "TU-1199, KEK-QUP-2023-0015",
    doi = "10.1016/j.physletb.2023.138213",
    journal = "Phys. Lett. B",
    volume = "846",
    pages = "138213",
    year = "2023"
}

@article{Jiang:2023qbm,
    author = "Jiang, Siyu and Yang, Aidi and Ma, Jiucheng and Huang, Fa Peng",
    title = "{Implication of nano-Hertz stochastic gravitational wave on dynamical dark matter through a dark first-order phase transition}",
    eprint = "2306.17827",
    archivePrefix = "arXiv",
    primaryClass = "hep-ph",
    doi = "10.1088/1361-6382/ad24c6",
    journal = "Class. Quant. Grav.",
    volume = "41",
    number = "6",
    pages = "065009",
    year = "2024"
}

@article{Eichhorn:2023gat,
    author = "Eichhorn, Astrid and Lino dos Santos, Rafael R. and Miqueleto, Jo{\~a}o Lucas",
    title = "{From quantum gravity to gravitational waves through cosmic strings}",
    eprint = "2306.17718",
    archivePrefix = "arXiv",
    primaryClass = "gr-qc",
    reportNumber = "MS-TP-23-27",
    doi = "10.1103/PhysRevD.109.026013",
    journal = "Phys. Rev. D",
    volume = "109",
    number = "2",
    pages = "026013",
    year = "2024"
}

@article{Cai:2023dls,
    author = "Cai, Yi-Fu and He, Xin-Chen and Ma, Xiao-Han and Yan, Sheng-Feng and Yuan, Guan-Wen",
    title = "{Limits on scalar-induced gravitational waves from the stochastic background by pulsar timing array observations}",
    eprint = "2306.17822",
    archivePrefix = "arXiv",
    primaryClass = "gr-qc",
    doi = "10.1016/j.scib.2023.10.027",
    journal = "Sci. Bull.",
    volume = "68",
    pages = "2929--2935",
    year = "2023"
}

@article{Lazarides:2023ksx,
    author = "Lazarides, George and Maji, Rinku and Shafi, Qaisar",
    title = "{Superheavy quasistable strings and walls bounded by strings in the light of NANOGrav 15~year data}",
    eprint = "2306.17788",
    archivePrefix = "arXiv",
    primaryClass = "hep-ph",
    doi = "10.1103/PhysRevD.108.095041",
    journal = "Phys. Rev. D",
    volume = "108",
    number = "9",
    pages = "095041",
    year = "2023"
}

@article{Blasi:2023sej,
    author = {Blasi, Simone and Mariotti, Alberto and Rase, A{\"a}ron and Sevrin, Alexander},
    title = "{Axionic domain walls at Pulsar Timing Arrays: QCD bias and particle friction}",
    eprint = "2306.17830",
    archivePrefix = "arXiv",
    primaryClass = "hep-ph",
    doi = "10.1007/JHEP11(2023)169",
    journal = "J. High Energy Phys.",
    volume = "11",
    pages = "169",
    year = "2023",
    number = "2023"
}

@article{Han:2023olf,
    author = "Han, Chengcheng and Xie, Ke-Pan and Yang, Jin Min and Zhang, Mengchao",
    title = "{Self-interacting dark matter implied by nano-Hertz gravitational waves}",
    eprint = "2306.16966",
    archivePrefix = "arXiv",
    primaryClass = "hep-ph",
    doi = "10.1103/PhysRevD.109.115025",
    journal = "Phys. Rev. D",
    volume = "109",
    number = "11",
    pages = "115025",
    year = "2024"
}

@article{Gouttenoire:2023ftk,
    author = "Gouttenoire, Yann and Vitagliano, Edoardo",
    title = "{Domain wall interpretation of the PTA signal confronting black hole overproduction}",
    eprint = "2306.17841",
    archivePrefix = "arXiv",
    primaryClass = "gr-qc",
    doi = "10.1103/PhysRevD.110.L061306",
    journal = "Phys. Rev. D",
    volume = "110",
    number = "6",
    pages = "L061306",
    year = "2024"
}

@article{Schwaller:2015tja,
    author = "Schwaller, Pedro",
    title = "{Gravitational Waves from a Dark Phase Transition}",
    eprint = "1504.07263",
    archivePrefix = "arXiv",
    primaryClass = "hep-ph",
    reportNumber = "CERN-PH-TH-2015-093",
    doi = "10.1103/PhysRevLett.115.181101",
    journal = "Phys. Rev. Lett.",
    volume = "115",
    number = "18",
    pages = "181101",
    year = "2015"
}

@article{Breitbach:2018ddu,
    author = "Breitbach, Moritz and Kopp, Joachim and Madge, Eric and Opferkuch, Toby and Schwaller, Pedro",
    title = "{Dark, Cold, and Noisy: Constraining Secluded Hidden Sectors with Gravitational Waves}",
    eprint = "1811.11175",
    archivePrefix = "arXiv",
    primaryClass = "hep-ph",
    reportNumber = "CERN-TH-2018-255, MITP/18-115",
    doi = "10.1088/1475-7516/2019/07/007",
    journal = "J. Cosmol. Astropart. Phys.",
    volume = "07",
    pages = "007",
    year = "2019",
    number = "2019"
}

@article{Fairbairn:2019xog,
    author = "Fairbairn, Malcolm and Hardy, Edward and Wickens, Alastair",
    title = "{Hearing without seeing: gravitational waves from hot and cold hidden sectors}",
    eprint = "1901.11038",
    archivePrefix = "arXiv",
    primaryClass = "hep-ph",
    reportNumber = "KCL-PH-TH/2019-12",
    doi = "10.1007/JHEP07(2019)044",
    journal = "J. High Energy Phys.",
    volume = "07",
    pages = "044",
    year = "2019",
    number = "2019"
}

@article{Caprini:2024gyk,
    author = "Caprini, Chiara and Jinno, Ryusuke and Konstandin, Thomas and Roper Pol, Alberto and Rubira, Henrique and Stomberg, Isak",
    title = "{Gravitational waves from first-order phase transitions: from weak to strong}",
    eprint = "2409.03651",
    archivePrefix = "arXiv",
    primaryClass = "gr-qc",
    doi = "10.1007/JHEP07(2025)217",
    journal = "J. High Energy Phys.",
    volume = "07",
    pages = "217",
    year = "2025",
    number = "2025"
}

@article{Correia:2025qif,
    author = "Correia, Jos{\'e} and Hindmarsh, Mark and Rummukainen, Kari and Weir, David J.",
    title = "{Gravitational waves from strong first order phase transitions}",
    eprint = "2505.17824",
    archivePrefix = "arXiv",
    primaryClass = "astro-ph.CO",
    month = "5",
    year = "2025"
}

@article{Chala:2025cya,
    author = "Chala, Mikael and Dashko, Andrii and Guedes, Guilherme",
    title = "{Running Couplings in High-Temperature Effective Field Theory}",
    eprint = "2510.26878",
    archivePrefix = "arXiv",
    primaryClass = "hep-ph",
    reportNumber = "CERN-TH-2025-220",
    month = "10",
    year = "2025"
}

@article{Chala:2024xll,
    author = "Chala, Mikael and Criado, Juan Carlos and Gil, Luis and Miras, Javier L{\'o}pez",
    title = "{Higher-order-operator corrections to phase-transition parameters in dimensional reduction}",
    eprint = "2406.02667",
    archivePrefix = "arXiv",
    primaryClass = "hep-ph",
    doi = "10.1007/JHEP10(2024)025",
    journal = "J. High Energy Phys.",
    volume = "10",
    pages = "025",
    year = "2024",
    number = "2024"
}

@article{Chala:2025oul,
    author = "Chala, Mikael and Gil, Luis and Ren, Zhe",
    title = "{Phase transitions in dimensional reduction up to three loops*}",
    eprint = "2505.14335",
    archivePrefix = "arXiv",
    primaryClass = "hep-ph",
    doi = "10.1088/1674-1137/adf322",
    journal = "Chin. Phys.",
    volume = "49",
    number = "12",
    pages = "123105",
    year = "2025"
}

@article{Bernardo:2025vkz,
    author = "Bernardo, Fabio and Klose, Philipp and Schicho, Philipp and Tenkanen, Tuomas V. I.",
    title = "{Higher-dimensional operators at finite temperature affect gravitational-wave predictions}",
    eprint = "2503.18904",
    archivePrefix = "arXiv",
    primaryClass = "hep-ph",
    reportNumber = "HIP-2025-6/TH",
    doi = "10.1007/JHEP08(2025)109",
    journal = "J. High Energy Phys.",
    volume = "08",
    pages = "109",
    year = "2025",
    number = "2025"
}

@article{Navarrete:2025yxy,
    author = {Navarrete, Pablo and Paatelainen, Risto and Sepp{\"a}nen, Kaapo and Tenkanen, Tuomas V. I.},
    title = "{Cosmological phase transitions without high-temperature expansions}",
    eprint = "2507.07014",
    archivePrefix = "arXiv",
    primaryClass = "hep-ph",
    reportNumber = "HIP-2025-20/TH",
    month = "7",
    year = "2025"
}

@article{Chakrabortty:2024wto,
    author = "Chakrabortty, Joydeep and Mohanty, Subhendra",
    title = "{One Loop Thermal Effective Action}",
    eprint = "2411.14146",
    archivePrefix = "arXiv",
    primaryClass = "hep-th",
    doi = "10.1016/j.nuclphysb.2025.117165",
    journal = "Nucl. Phys. B",
    volume = "1020",
    pages = "117165",
    year = "2025"
}

@article{Balui:2025yvd,
    author = "Balui, Debanjan and Biswas, Tisa and Chakrabortty, Joydeep and Dey, Debmalya and Englert, Christoph and Mohanty, Subhendra",
    title = "{Gauge choices, infrared pitfalls, and thermal effects in effective potentials}",
    eprint = "2507.22706",
    archivePrefix = "arXiv",
    primaryClass = "hep-th",
    doi = "10.1103/drsd-wfns",
    journal = "Phys. Rev. D",
    volume = "112",
    number = "5",
    pages = "056022",
    year = "2025"
}

@article{Vassilevich:2003xt,
    author = "Vassilevich, D. V.",
    title = "{Heat kernel expansion: User's manual}",
    eprint = "hep-th/0306138",
    archivePrefix = "arXiv",
    doi = "10.1016/j.physrep.2003.09.002",
    journal = "Phys. Rept.",
    volume = "388",
    pages = "279--360",
    year = "2003"
}

@article{Yamada:2025hfs,
    author = "Yamada, Masaki",
    title = "{Maximal GW amplitude from bubble collisions in supercooled phase transitions}",
    eprint = "2509.13402",
    archivePrefix = "arXiv",
    primaryClass = "gr-qc",
    reportNumber = "TU-1275",
    month = "9",
    year = "2025"
}

@article{Yamada:2025cfr,
    author = "Yamada, Masaki",
    title = "{Analytic derivation of GW spectrum from bubble collisions in FLRW Universe}",
    eprint = "2509.16073",
    archivePrefix = "arXiv",
    primaryClass = "astro-ph.CO",
    reportNumber = "TU-1276",
    month = "9",
    year = "2025"
}

@article{Kierkla:2025vwp,
    author = "Kierkla, Maciej and Ramberg, Nicklas and Schicho, Philipp and Schmitt, Daniel",
    title = "{Theoretical uncertainties for primordial black holes from cosmological phase transitions}",
    eprint = "2506.15496",
    archivePrefix = "arXiv",
    primaryClass = "hep-ph",
    reportNumber = "SISSA 07/2025/FISI",
    month = "6",
    year = "2025"
}
\end{document}